\documentclass[useAMS,usenatbib]{mn2e}

\usepackage{amsmath}
\usepackage{amssymb}
\usepackage{graphicx}
\usepackage{txfonts}
\usepackage{natbib}
\graphicspath {{./figures/}}

\newcommand{\dd}{\mathrm{d}}
\newcommand{\e}{\mathrm{e}}

\newcommand{\aap}{A\&A}

\newcommand{\ao}{Appl. Opt.}
\newcommand{\apj}{ApJ}
\newcommand{\apjs}{ApJS}
\newcommand{\apjl}{ApJL}
\newcommand{\araa}{ARA\&A}
\newcommand{\mnras}{MNRAS}
\newcommand{\pasp}{Publi. Astron. Soc. Pac.}
\newcommand{\prd}{Phys. Rev. D}
\newcommand{\jcap}{J. Cosmology Astroparticle Phys.}

\title[A GEV application to massive galaxy clusters]{An application of extreme value statistics to the most massive galaxy clusters at low and high redshifts}

\author[J.-C. Waizmann, S. Ettori and L. Moscardini]{J.-C. Waizmann$^{1,2}$\thanks{E-mail:
jcwaizmann@oabo.inaf.it}, S. Ettori$^{1,2}$ and L. Moscardini$^{3,1,2}$\\
$^{1}$INAF - Osservatorio Astronomico di Bologna, via Ranzani 1, 40127 Bologna, Italy\\
$^{2}$INFN, Sezione di Bologna, viale Berti Pichat 6/2, 40127 Bologna, Italy \\
$^{3}$Dipartimento di Astronomia, Università di Bologna, via Ranzani 1, 40127 Bologna, Italy}
\begin{document}

\date{Accepted 2011 November 9. Received 2011 November 8; in original form 2011 September 22}

\pagerange{\pageref{firstpage}--\pageref{lastpage}} \pubyear{2011}

\maketitle

\label{firstpage}

\begin{abstract}
In this work we present an application of general extreme value statistics (GEV) to very massive single clusters at high and low redshifts. After introducing the formalism, we apply this statistics to four very massive high redshift clusters. Those clusters comprise ACT-CL~J0102-4915 with a mass of $M_{200\rm m}=(2.16\pm 0.32)\times 10^{15}\,M_\odot$ at a redshift of $z=0.87$, SPT-CL~J2106-5844 with a mass of $M_{200\rm m}=(1.27\pm 0.21)\times 10^{15}\,M_\odot$ at $z=1.132$ and two clusters found by the \textit{XMM-Newton} Distant Cluster Project survey: XMMU~J2235.32557 with a mass of $M_{200\rm c}= (7.3 \pm 1.3) \times 10^{14}\,M_\odot$ located at a redshift $z=1.4$ and XMMU~J0044.0-2033 having a mass in the range of $M_{200\rm c}= (3.5-5.0) \times 10^{14}\,M_\odot$ at $z=1.579$. By relating those systems to their corresponding distribution functions of being the most massive system in a given survey area and redshift interval, we find that none of the systems alone is in tension with $\Lambda$ cold dark matter ($\Lambda$CDM). We confront these results with a GEV analysis of four very massive low redshift clusters: A2163, A370, RXJ1347-1145 and 1E0657-558, finding no tendency of the high-$z$ systems to be more extreme than the low-$z$ ones. \\
In addition, we study the extreme quantiles of single clusters at high-$z$ and present contour plots for fixed quantiles in the mass vs. survey area plane for four redshift intervals, finding that, in order to be significantly in conflict with $\Lambda$CDM, cluster masses would have to be substantially higher than the currently observed ones.

\end{abstract}

\begin{keywords}
methods: statistical -- galaxies: clusters: individual (ACT- CL J0102-4915, SPT-CL J2106-5844, XMMU J0044.0-2033) -- galaxies: clusters: general -- cosmology: observations  
\end{keywords}

%------------------------------------------
\section{Introduction}\label{sec:intro}
%------------------------------------------
The discovery of the high-redshift cluster XMMU~J2235.3−2557 at $z=1.4$ by \cite{Mullis2005} and its following joint X-ray and lensing analysis \citep{Rosati2009, Jee2009}, which established this system as the most massive cluster at redshift $z>1$ at the time, have motivated a number of studies about the usability of very massive galaxy clusters at high redshifts as a cosmological probe \citep{Holz2010, Hoyle2011, Mantz2008, Mantz2010, Mortonson2011, Hotchkiss2011, Hoyle2011b}. The majority of those mainly focus on consistency-checks of the $\Lambda$ cold dark matter ($\Lambda$CDM) concordance model, some discuss alternatives to the standard model in the form of coupled dark energy \citep{Baldi2011a, Baldi2011b} or non-Gaussianity \citep{Cayon2011, Enqvist2011, Paranjape2011, Chongchitnan2011}.\\
In parallel to these developments, the application of extreme value theory to high-mass clusters at high redshifts became increasingly popular. \cite{Davis2011} utilised general extreme value statistics (GEV) (see e.g. \cite{Gumbel1958, Kotz2000, Coles2001}) in order to study the probability distribution of the most massive halo in a given volume. In addition, \cite{Colombi2011} applied GEV to the statistics of Gaussian random fields. Based on this groundwork, \cite{Waizmann2011} proposed to use GEV for reconstructing the cumulative distribution function of massive clusters from high-$z$ cluster surveys and to use it as a discriminant between different cosmological models. Recently, \cite{Harrison&Coles2011} also studied extreme value statistics based on the exact form rather than the asymptotic one. Other applications in the framework of astrophysics are the study of the statistics of the brightest cluster galaxies by \cite{Bhavsar1985} and the application to temperature maxima in the cosmic microwave background by \cite{Coles1988}.\\

In this work we present an application of extreme value statistics to individual massive clusters at high and low redshifts by computing the probability distribution functions for these clusters. We study the probability to find such clusters as the most massive systems also in ill defined survey areas and compare our results to the findings of other works. Furthermore, we utilise GEV for the computation of extreme quantiles in order to quickly relate an arbitrarily observed cluster to its probability of occurrence. \\

This paper is structured according to the following scheme. In Section~\ref{sec:GEV}, we briefly introduce the application of GEV to massive clusters as discussed by \cite{Davis2011}. An introduction of the clusters studied in this work follows in Section~\ref{sec:GEVclust} and it is divided into subsections for the high-$z$ and low-$z$ systems. We apply GEV to the chosen objects in Section~\ref{sec:gev_analysis}: in Section~\ref{subsec:bias} we discuss the bias arising from a posteriori choice of the redshift intervals for the analysis and in Section~\ref{subsec:mass_corr} how the observed cluster mass has to be corrected for a subsequent statistical analysis. This analysis is then performed in Section~\ref{subsec:high_z} for the four most massive clusters in four a priori defined redshift intervals. In Section~\ref{sec:lowVShigh}, we study the impact of the survey area on the existence probabilities and compare the low with the high-$z$ clusters. In the following Section~\ref{sec:quantiles}, we introduce the extreme quantiles based on GEV and compute the iso-quantile contours for four redshift intervals in the mass-survey area plane. After this, we summarise our findings in the conclusions in Section~\ref{sec:conclusions}.
%------------------------------------------------------
\section{GEV statistics in a cosmological context}\label{sec:GEV}
%------------------------------------------------------
Extreme value theory (for an introduction see e.g. \cite{Gumbel1958, Kotz2000, Coles2001}) is concerned with the stochastic behaviour of the maxima or minima of i.i.d. random variables $X_i$. In what follows we will only consider the first case, introducing the block maximum $M_n$ defined as
\begin{equation}
M_n=\max(X_1,\dotsc X_n).
\label{eq:blockmax}
\end{equation}
It has been shown \citep{Fisher1928, Gnedenko1943} that, for $n\rightarrow\infty$, the limiting cumulative distribution function (CDF) of the renormalised block maxima is given by one of the extreme value families: Gumbel (Type I), Fr\'{e}chet (Type II) or Weibull (Type III). As independently shown by \cite{vonmises1954} and \cite{Jenkinson1955}, these three families can be unified as a general extreme value distribution (GEV)
\begin{equation}\label{eq:GEV}
  G_{\gamma,\,\beta,\,\alpha}(x) = \left\{ 
  \begin{array}{l l}
    \exp{\left\lbrace -\left[1+\gamma \left(\frac{x-\alpha}{\beta}\right)\right]^{-1/\gamma}\right\rbrace}, & \quad {\rm for}\quad\gamma\neq 0,\\
    \exp{\left\lbrace \e^{-\left(\frac{x-\alpha}{\beta}\right)}\right\rbrace} ,& \quad {\rm for}\quad\gamma = 0,\\
  \end{array} \right.
\end{equation}
with the shape-, scale- and location parameters $\gamma$, $\beta$ and $\alpha$. In this generalisation, $\gamma=0$ corresponds to the Type I,  $\gamma>0$ to Type II and $\gamma<0$ to the Type III distributions. The corresponding probability density function (PDF) is given by
\begin{equation}
g_{\gamma,\,\beta,\,\alpha}(x)=\frac{\dd G_{\gamma,\,\beta,\,\alpha}(x)}{\dd x}.
\end{equation}
From now on we will adopt the convention that capital initial letters denote the CDF (like $G_{\gamma,\,\beta,\,\alpha}(x)$) and small initial letters denote the PDF (like $g_{\gamma,\,\beta,\,\alpha}(x)$).\\

%-------------------------------------
\begin{figure}
\centering
\includegraphics[width=0.9\linewidth]{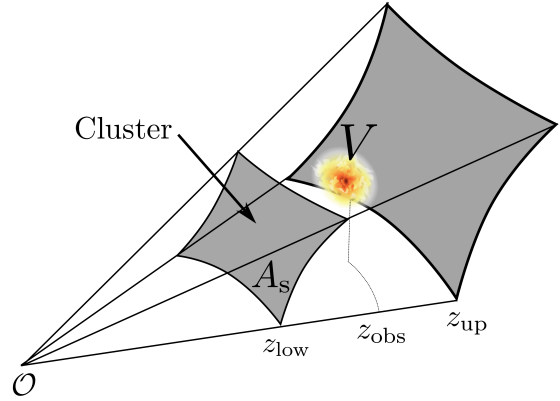}
\caption{Illustrative scheme for the application of GEV on a single cluster observed at $z_{\rm obs}$ in order to study its probability of existence in a given volume $V$. The volume is defined by the survey area, $A_{\rm s}$, and the redshift interval, $z\in [z_{\rm low}, z_{\rm up}]$.}\label{fig:scheme}
\end{figure}
%-------------------------------------
A formalism for the application of GEV on the most massive galaxy clusters has been introduced by \cite{Davis2011} and is briefly summarised in the following. By introducing the random variable $u\equiv\log_{10}(m)$, the CDF of the most massive halo reads
\begin{equation}
{\rm Pr}\lbrace u_{\rm max} \le u\rbrace
\equiv \int_0^{u}p(u_{\rm max}) \,\dd u_{\rm max}.
\label{eq:cdf}
\end{equation}
This probability has to be equal to the one of finding no halo with a mass larger than $u$. On scales  ($\geq 100\,\textrm{Mpc}\,h^{-1}$), for which the clustering between galaxy clusters can be neglected, the CDF is given by the Poisson distribution for the case of zero occurence \citep{Davis2011}:
\begin{equation}
P_0(u)=\frac{\lambda^k \exp{\left(-\lambda\right)}}{k!}=\exp\left[-n_{\rm eff}( > u) V\right],
\end{equation}
where $n_{\rm eff}( > u)$ is the effective comoving number density of halos above mass $u=\log_{10}(m)$ obtained by averaging and $V$ is the comoving volume. By assuming that equation~\eqref{eq:cdf} can be modelled by $G_{\gamma,\,\beta,\,\alpha}(u)$, it is possible to relate the GEV parameters to cosmological quantities by Taylor-expanding both $G_{\gamma,\,\beta,\,\alpha}(u)$ and $P_0(u)$ around the peaks of the corresponding PDFs: 
\begin{align*}
 P_0(u)&=P_0(u_0)+\left.\frac{\dd\,P_0(u)}{\dd\,u}\right|_{u_0} \left(u-u_0\right)+\dots\; ,\\
G_{\gamma,\,\beta,\,\alpha}(u)&=G_{\gamma,\,\beta,\,\alpha}(u_0)+\left.\frac{\dd\,G_{\gamma,\,\beta,\,\alpha}(u)}{\dd\,u}\right|_{u_0} \left(u-u_0\right)+\dots\; .
\end{align*}
By comparing the individual first two expansion terms with each other, one finds \citep{Davis2011} 
\begin{eqnarray}\label{eq:parameters}
\gamma = n_{\rm eff}(>m_0)V-1, \quad \beta =
\frac{(1+\gamma)^{(1+\gamma)}}{\left.\frac{\dd\,n_{\rm eff}}{\dd\,m}\right|_{m_0}Vm_0\ln 10}, \nonumber \\
\alpha = \log_{10} m_0 - \frac{\beta}{\gamma}[(1+\gamma)^{-\gamma} -1],
\end{eqnarray}
where $m_0$ is the most likely maximum mass and $\left.\dd\,n_{\rm eff} / \dd\,m \right|_{m_0}$ is the effective mass function evaluated at $m_0$, which relates to the effective number density $n_{\rm eff}(>m)$ via
 \begin{equation}
 \left.\frac{\dd\,n_{\rm eff}}{\dd\,m}\right|_{m_0}=- \left.\frac{\dd\,n_{\rm eff}(>m)}{\dd\,m}\right|_{m_0}.
\end{equation}
 The most likely mass, $m_0$, can be found \citep{Davis2011, Waizmann2011} by performing a root search on
\begin{equation}\label{eq:m0_num}
 \left.\frac{\dd\,n_{\rm eff}}{\dd\,m}\right|_{m_0}+m_0\left.\frac{\dd ^2\,n_{\rm eff}}{\dd\,m^2}\right|_{m_0}+m_0V\left(\left.\frac{\dd\,n_{\rm eff}}{\dd\,m}\right|_{m_0}\right)^2=0,
\end{equation}
For calculating $n_{\rm eff}$ we utilised the mass function introduced by \cite{Tinker2008} and fix the cosmology to $(h,\Omega_{\Lambda0},\Omega_{\rm m0},\sigma_8)=(0.7,0.73,0.27,0.81)$ based on the \textit{Wilkinson Microwave Anisotropy} 7-yr (\textit{WMAP}7) results \citep{Komatsu2011}.\\
As illustrated in Fig.~\ref{fig:scheme}, the volume $V$ is determined by the survey area and the redshift interval $z\in [z_{\rm low}, z_{\rm up}]$, where $z_{\rm low}$ and $z_{\rm up}$ are the lower and upper boundary, respectively, which contain the individual cluster at the observed redshift $z_{\rm obs}$. The choice of the individual redshift intervals is discussed in further detail in Section~\ref{sec:gev_analysis}.
%-------------------------------------
\begin{figure*}
\centering
\includegraphics[width=0.465\linewidth]{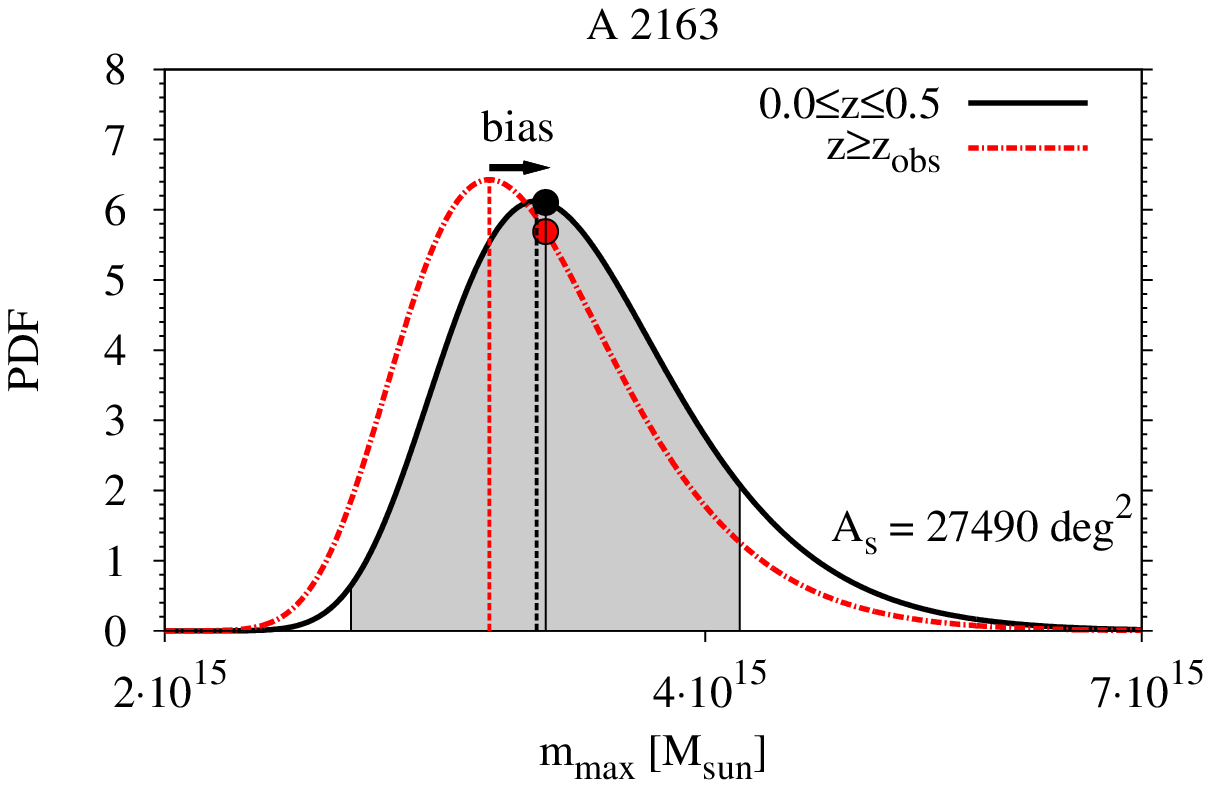}
\includegraphics[width=0.465\linewidth]{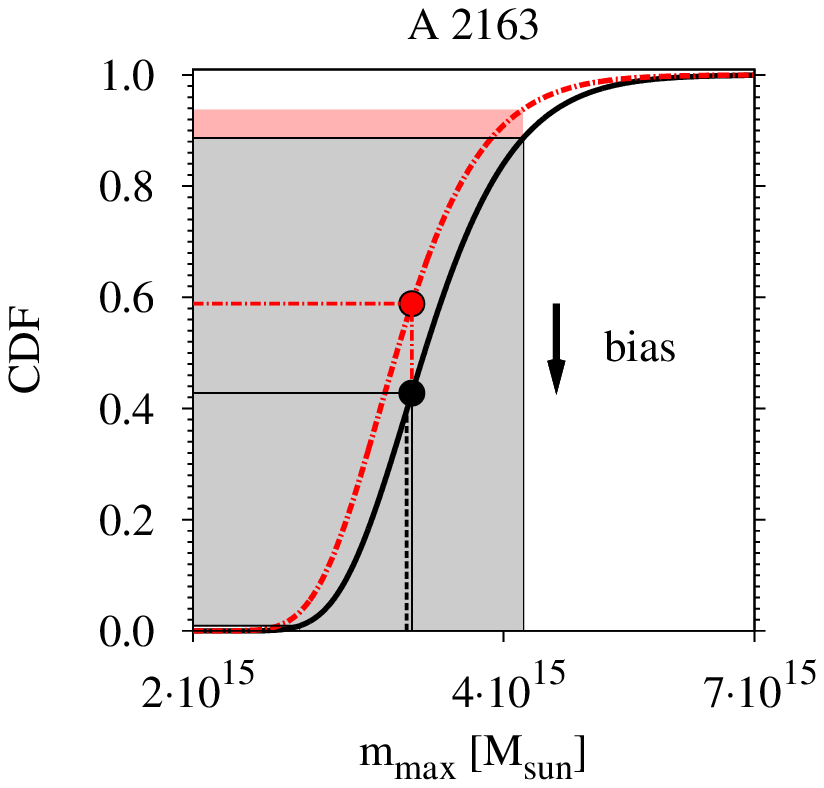}\\
\includegraphics[width=0.465\linewidth]{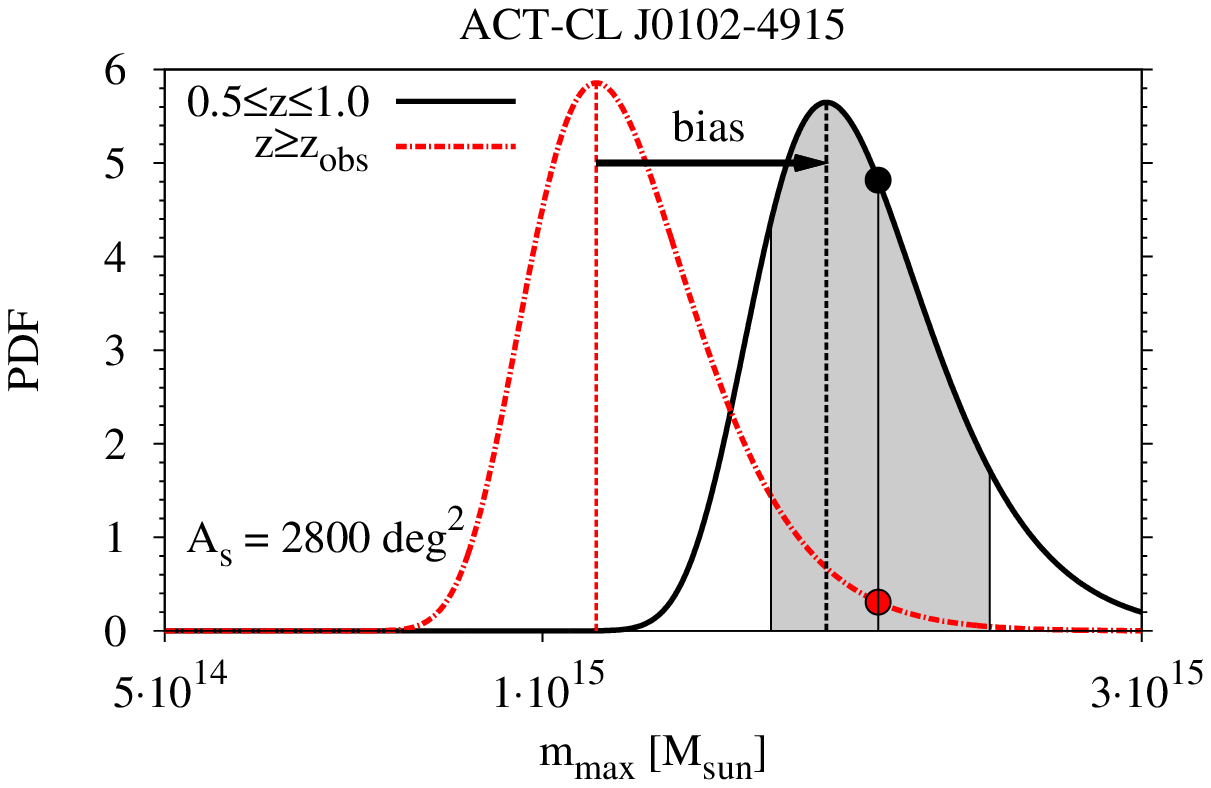}
\includegraphics[width=0.465\linewidth]{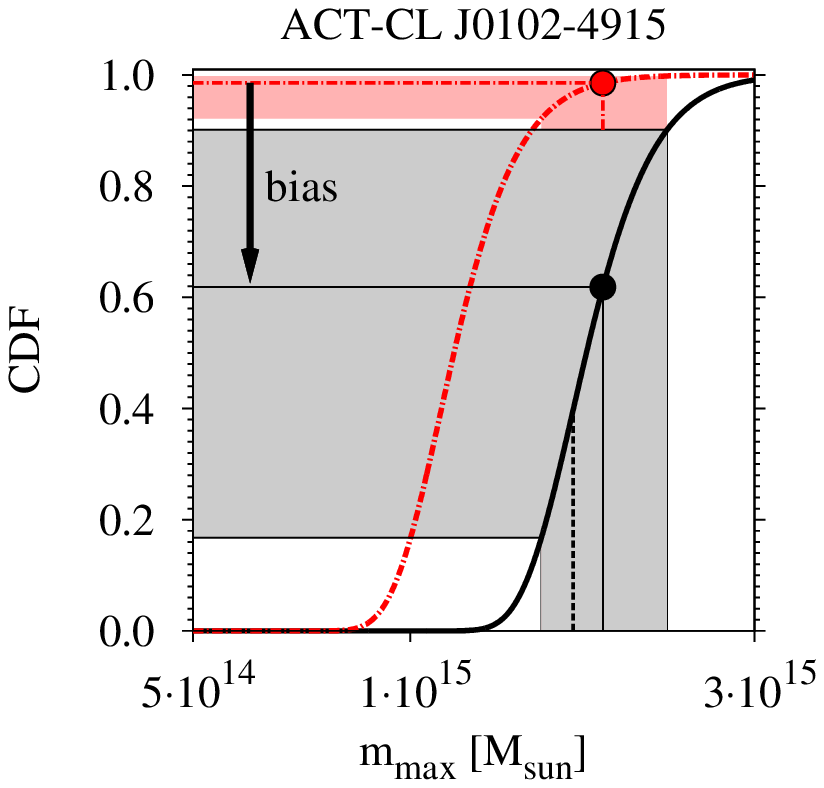}\\
\includegraphics[width=0.465\linewidth]{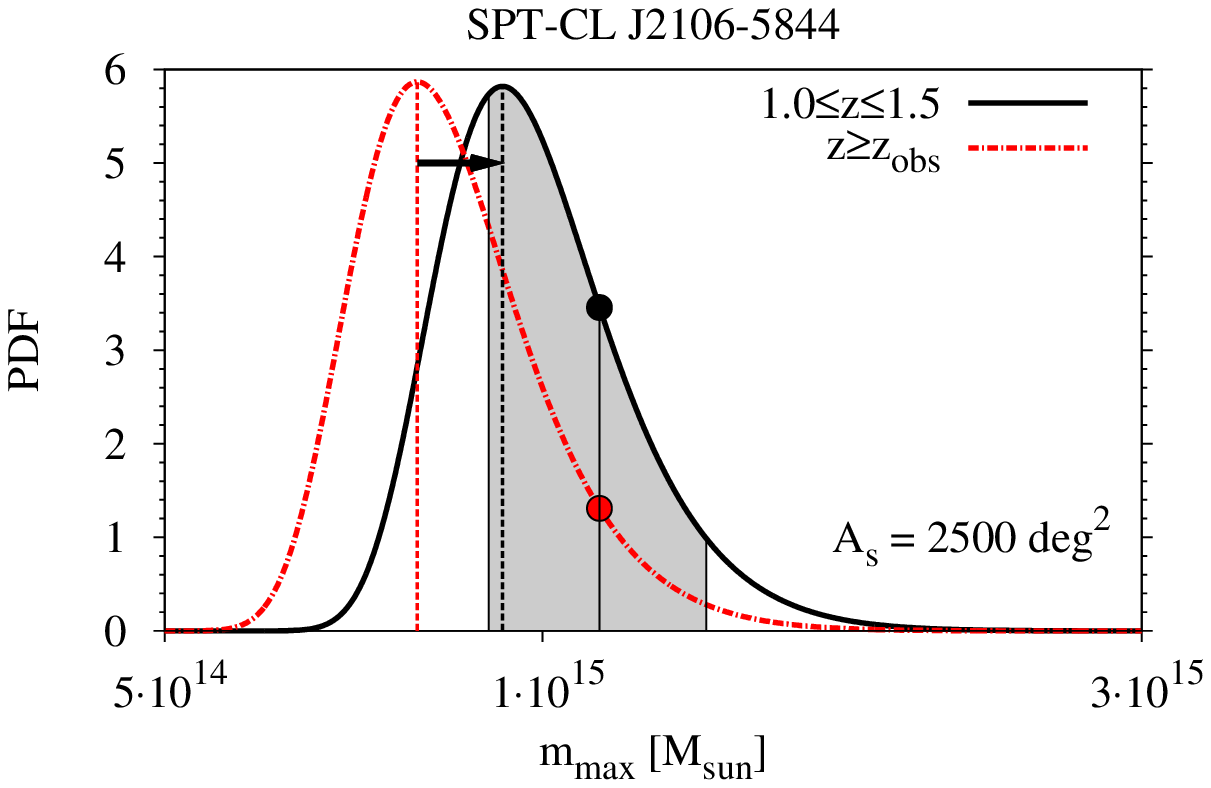}
\includegraphics[width=0.465\linewidth]{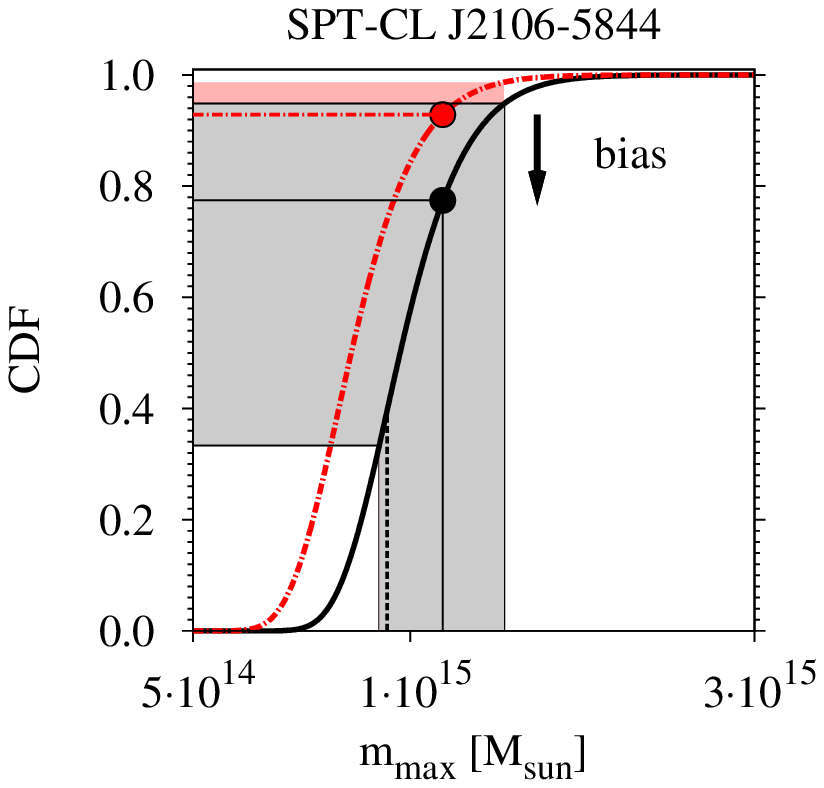}\\
\includegraphics[width=0.465\linewidth]{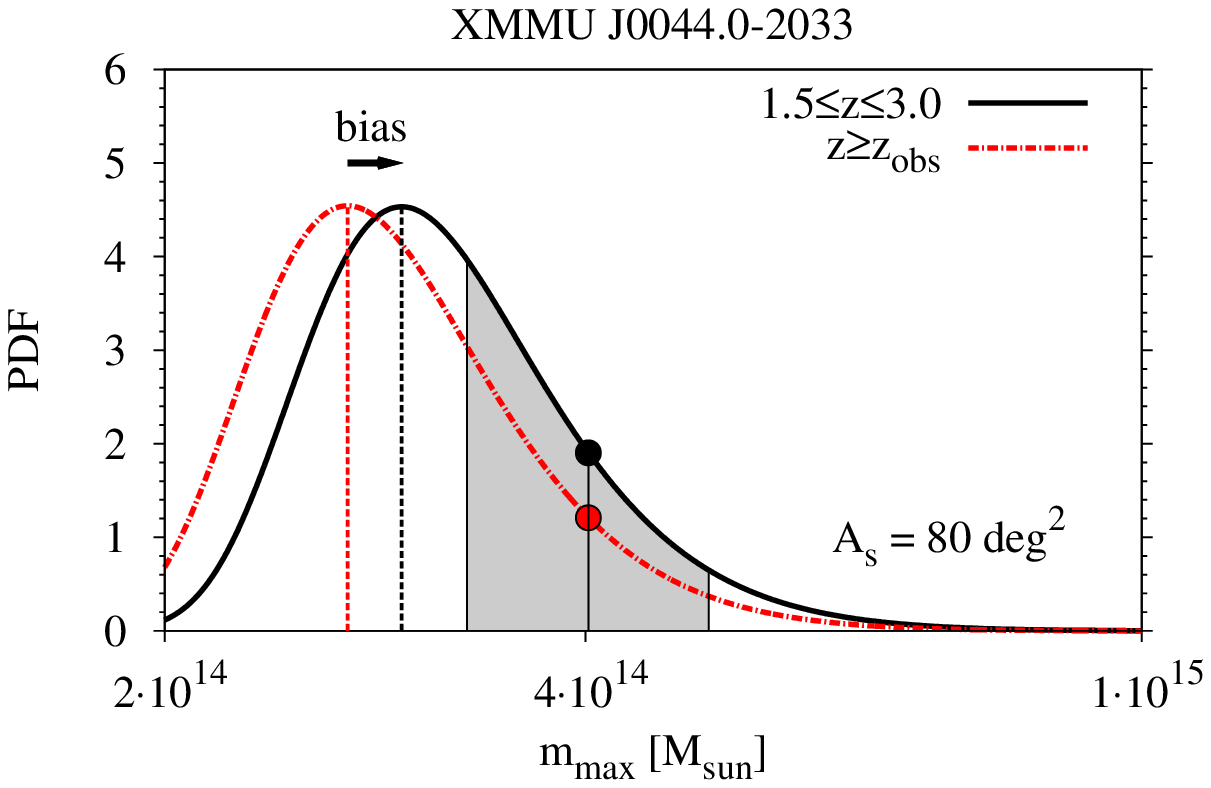}
\includegraphics[width=0.465\linewidth]{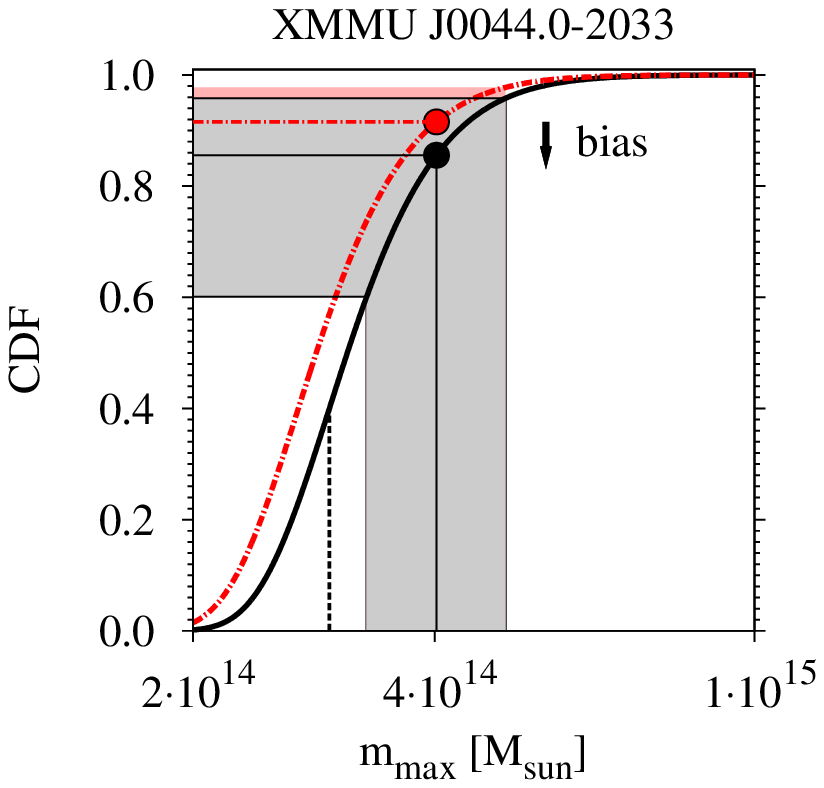}
\caption{PDFs (left-hand column) and CDFs (right-hand column) for the four most massive clusters in the redshift bins given in Tab.~\ref{tab:bins}. For all four clusters the effect of the bias discussed in Section~\ref{sec:gev_analysis} is shown, where the red, dash-dotted lines are the distributions assuming the redshift interval $z\in [z_{\rm obs}, z_{\rm up}]$ and the black, solid lines for the $z$-interval given in the key (left-hand panels). The Eddington bias corrected position of the system on the distributions is given by the filled black (red) circle for both cases, illustrating the impact of the aforementioned bias. The grey (red) shaded regions represent the regions of uncertainty in the mass measurements and the central line in these regions refers to the reported mass of the cluster. The dashed lines denote the most likely expected mass of the most massive system for the given survey area and redshift interval.}\label{fig:spt}
\end{figure*}
%-------------------------------------
%-------------------------------------
\begin{table*}
% \begin{minipage}{126mm}
\caption{Compilation of the data of the galaxy clusters studied in this work. The masses $M_{200\rm c}$ and $M_{200\rm m}$ are with respect to the critical and the mean background density. The mass $M_{200\rm m}^{\rm Edd}$ gives the mass after the correction for the Eddington bias based on the estimated mass uncertainty $\sigma_{\ln M}$. The last column lists the references for the observed mass (either $M_{200\rm c}$, or $M_{200\rm m}$), on which the analysis is based on.}
\begin{tabular}{lcccccc} \hline
Cluster & $z$ & $M_{200\rm c}$ in units of $M_\odot$ & $M_{200\rm m}$ in units of $M_\odot$ & $\sigma_{\ln M}$ & $M_{200\rm m}^{\rm Edd}$ in units of $M_\odot$ & Reference \\
\hline
ACT-CL~J0102  & $0.87$ & $-$ & $(2.16\pm 0.32)\times 10^{15}$ & $0.2$ & $1.85_{-0.33}^{+0.42}\times 10^{15}$ & \cite{Marriage2011}  \\
SPT-CL~J2106  & $1.132$ & $-$ & $(1.27\pm 0.21)\times 10^{15}$ & $0.2$ & $1.11_{-0.20}^{+0.24}\times 10^{15}$ & \cite{Foley2011} \\
XXMU~J2235  & $1.4$ & $(7.3\pm 1.3)\times 10^{14}$ & $(7.74\pm1.38)\times 10^{14}$ & $0.2$ & $6.82_{-1.23}^{+1.52}\times 10^{14}$ & \cite{Jee2009}  \\
XXMU~J0044  & $1.579$ & $(4.25\pm 0.75)\times 10^{14}$ & $(4.46\pm0.79)\times 10^{14}$ & $0.3$ & $4.02_{-0.73}^{+0.88}\times 10^{14}$ & \cite{Santos2011} \\ \\
%-----------------------------------------------
A2163  & $0.203$ & $(2.7\pm 0.6)\times 10^{15}$ & $(3.68\pm0.82)\times 10^{15}$ & $0.25$ & $3.04^{+0.87}_{-0.67}\times 10^{15}$ & \cite{Maughan2011} \\
1E0657-558  & $0.296$ & $(1.75\pm 0.29)\times 10^{15}$ & $(2.28\pm0.38)\times 10^{15}$ & $0.2$ & $2.06^{+0.46}_{-0.37}\times 10^{15}$ & \cite{Maughan2011} \\
  &  & $-$ & $(3.12\pm1.15)\times 10^{15}$ & $0.45$ & $1.70_{-0.62}^{+0.96}\times 10^{15}$ & \cite{Williamson2011} \\
A370  & $0.375$ & $(2.21\pm 0.27)\times 10^{15}$ & $(2.79\pm 0.34)\times 10^{15}$ & $0.15$ & $2.61^{+0.42}_{-0.37}\times 10^{15}$  & \cite{Umetsu2011} \\
RXJ1347 & $0.451$ & $(2.1\pm 0.5)\times 10^{15}$ & $(2.59\pm0.62)\times 10^{15}$ & $0.25$ & $2.14^{+0.60}_{-0.48}\times 10^{15}$ & \cite{Maughan2011} \\ 
\hline
\end{tabular}\label{tab:clusters}
%\end{minipage} 
\end{table*}
%-------------------------------------
%------------------------------------------------------
\section{The studied clusters}\label{sec:GEVclust}
%------------------------------------------------------
In order to demonstrate the usability of GEV for the study of very massive clusters at high redshifts, we decided to apply the method to several observed clusters known to be rather massive given their detection redshift. We divide our analysis into two parts: in the first one, we consider four high-$z$ systems and in the second one four systems in the low redshift Universe.
\subsection{The high-$z$ systems}
As high redshift systems, we chose the following four objects: 
\begin{description}
\item[\textbf{ACT-CL~J0102-4915:}] This recently discovered \citep{Marriage2011} merging system is currently the most massive cluster observed at $z>0.6$ \citep{Menanteau2011} and has been detected by the \textit{Atacama Cosmology Telescope} (\textit{ACT}) \citep{Fowler2007} in its $755\,{\rm deg}^2$ field. Its mass has been determined by a combination of Sunyaev-Zeldovich (SZ) \citep{Sunyaev1972, Sunyaev1980}, optical (\textit{Very Large Telescope}), X-ray (\textit{Chandra}), and infrared (\textit{Spitzer}) data to be $M_{200\rm m}=(2.16\pm 0.32)\times 10^{15}\,M_\odot$ at a spectroscopic redshift of $z=0.87$. Since the survey areas of \textit{ACT} and the \textit{South Pole Telescope} (\textit{SPT}) \citep{Carlstrom2011} overlap and this cluster lies in the overlap region, we conservatively decided to assign the combined survey area of $2\,800\,{\rm deg}^2$ to this system.
\item[\textbf{SPT-CL~J2106-5844:}]
This recently reported cluster \citep{Foley2011, Williamson2011} has been detected by the \textit{SPT} survey in its $2\,500\,{\rm deg}^2$ field. The mass inferred from a combination of SZ and X-ray information is found to be $M_{200\rm m}=(1.27\pm 0.21)\times 10^{15}\,M_\odot$ and spectroscopy of the member galaxies locates this cluster at a redshift of $z=1.132$. These observed parameters make this cluster the most massive cluster  at redshift $z>1$ and thus a highly interesting system to study.
\item[\textbf{XMMU J2235.32557:}]
This object has been found by the \textit{XMM-Newton Distant Cluster Project} (\textit{XDCP}) survey \citep{Mullis2005} in an area of $11\,{\rm deg}^2$. By combining X-ray, weak lensing and velocity dispersion measurements \citep{Mullis2005, Rosati2009, Jee2009} the mass of this cluster has been estimated to be $M_{200\rm c}= (7.3 \pm 1.3) \times 10^{14}\,M_\odot$ and it is located at a redshift $z=1.4$. XMMU~J2235 was the trigger of many studies on the possibility of using massive clusters at high redshifts as cosmological probes as already summarised in the Introduction.
\item[\textbf{XMMU J0044.0-2033:}]
Like XMMU~J2235, this cluster has recently been detected by the \textit{XDCP} survey \citep{Santos2011} in an area of $\sim 80\,{\rm deg}^2$. The cluster is located at $z=1.579$ and its mass was found to be in the range of $M_{200\rm c}= (3.5-5.0) \times 10^{14}\,M_\odot$, which puts this cluster into the forefront of currently known high-mass and high-$z$ clusters. For the current work we adopted the central value $M_{200\rm c}=(4.25\pm 0.75)\times 10^{14}\,M_\odot$ as the mass for this system. 
\end{description}
%-----------------------------------------
\subsection{The low-$z$ systems}
The chosen objects in the low redshift Universe comprise:
\begin{description}
\item[\textbf{Abell 2163:}]
The merging cluster A2163 is one of the hottest X-ray clusters on the sky \citep{Arnaud1992}. At a redshift of  $z=0.203$, this cluster has a mass of $M_{200\rm c}=(2.7\pm 0.6)\times 10^{15}\,M_\odot$ and has been studied at multiple wavelengths [see \cite{Squires1997, Radovich2008, Nord2009, Okabe2011}, for instance].
\item[\textbf{Abell 370:}] While RXJ1347 is known to be the most luminous X-ray cluster on the sky, A370 is known to be the most massive lensing cluster \citep{Broadhurst2008, Broadhurst&Barkana2008} with a mass of $M_{200\rm c}= (2.21\pm 0.27) \times 10^{15}\,M_\odot$ \citep{Umetsu2011} at a redshift of $z=0.375$.
\item[\textbf{RXJ1347-1145:}]
This object has been found by the \textit{ROSAT All Sky-Survey} (\textit{RASS}) and is known to be the most luminous X-ray cluster on the sky. It has been studied in the optical and X-ray (see e.g. \cite{Schindler1995, Schindler1997, Ettori2001}) as well as via the SZ \citep{Komatsu1999, Pointecouteau1999}. The mass of this cluster has been found to be $M_{200\rm c}= (2.1\pm 0.5) \times 10^{15}\,M_\odot$ and it is located at a redshift $z=0.451$. 
\item[\textbf{1E0657-558:}] Widely known as the "Bullet-Cluster" \citep{Tucker1998, Markevitch2002, Clowe2006}, this dynamically very interesting system is with $M_{200\rm m} =(3.12 \pm 1.15)\times 10^{15} M_{\odot}$\footnote{The statistical and systematic errors from Tab.~6 of \cite{Williamson2011} have been added in quadrature.} at $z=0.296$ the most massive cluster in the \textit{SPT} survey field \citep{Williamson2011} because of which it has been added to this study. In addition to the SZ based mass from above, we added also the X-ray mass to our study as discussed below.
\end{description}
The X-ray masses for A2163, RXJ1347 and 1E0657-558 have been estimated through the $M-T$ scaling relation in \cite{Arnaud2005} (Tab.~2), by adopting the X-ray temperature estimated in the $(0.15−1) R_{500}$ aperture from \cite{Maughan2011}: $15.2 \pm 1.2\,{\rm keV}$ for A2163, $14.2 \pm 1.4\,{\rm keV}$ for RXJ1347 and $11.7 \pm 0.5\,{\rm keV}$ for 1E0657 that correspond to $M_{200\rm c} =(2.7 \pm 0.6) \times 10^{15} M_{\odot}$, $(2.1 \pm 0.5) \times 10^{15} M_{\odot}$ and $(1.75 \pm 0.29) \times 10^{15} M_{\odot}$, respectively.
In order to apply a GEV analysis to the aforementioned low-$z$ clusters, we have to define the survey area. By adding the areas for the northern (\textit{NORAS}) and southern (\textit{REFLEX}) cluster samples of the \textit{RASS}, one finds $A_{\rm s}=27\,490\,{\rm deg}^2$, which we will adopt for our analysis. \\
It should be mentioned that all observed masses reported in the section are measured with respect to the critical background density, apart from the \textit{ACT} and \textit{SPT} clusters, whereas the masses from theory are usually defined with respect to the mean background density. This issue will be discussed in more detail in the following section. A compilation of the data of all studied clusters can be found in Tab.~\ref{tab:clusters}.
%-------------------------------------
\section{The GEV analysis}\label{sec:gev_analysis}
%-------------------------------------
In order to apply a GEV analysis to the aforementioned single systems with the goal of quantifying their probability of existence in a $\Lambda$CDM cosmology, one has has take several effects into account that will be discussed in the following. 
\subsection{Bias due to the a posteriori choice of the redshift interval}\label{subsec:bias}
The most important effect that has to be considered is the bias, introduced and discussed in detail in \cite{Hotchkiss2011}, that stems from the a posteriori definition of the redshift interval for which the likelihood of a cluster is calculated. When the redshift interval is defined a posteriori (see e.g. \cite{Mortonson2011} and \cite{Foley2011}), one ignores the fact that the potentially rare cluster of interest could have easily appeared at a different redshift leading to a different probability of existence. In order to avoid this bias we decided to define a priori four redshift bins $0\le z\le 0.5$,  $0.5\le z\le 1.0$,  $1.0\le z\le 1.5$ and $1.5\le z\le 3.0$ and to calculate the probability distributions of the most massive cluster in these intervals for a given survey area, $A_s$, as illustrated in Fig.~\ref{fig:scheme}. Then, we select, out of the clusters listed in Section~\ref{sec:GEVclust}, the most massive ones that fall in these individual bins and quantify their probability of existence. The effect of the bias is shown in Fig.~\ref{fig:spt} as a substantial shift of the probability distributions depending on the choice of the redshift interval and is discussed in more detail below in Section~\ref{subsec:high_z}. 
\subsection{Preparing the observed cluster masses for the analysis}\label{subsec:mass_corr}
Another obstacle to be overcome is the unification of the mass definitions. All observed masses $M_{200}$ stated above, apart from ACT-CL~J0102 and SPT-CL~J2106, are defined with respect to the \textit{critical} background density $\rho_{\rm c} (z) = 3 H^2(z) / (8 \pi G)$, whereas for the theoretical mass function the \textit{mean} background density is assumed $\rho_{\rm m}(z) = \Omega_{\rm m}(z) \rho_{\rm c}(z)$. In order to compare the two, we have to scale the cluster masses to the definition used in the mass function. For an arbitrary overdensity, $\Delta$, the mass is defined as 
\begin{equation}
M_{\Delta} = \frac{4}{3}\pi R_{\Delta}^3 \, \Delta \rho_{\rm c}(z), 
\label{eqa:mass}
\end{equation}
where $z$ is the cluster's redshift and $R_{\Delta} = c_{\Delta} r_{\rm s}$ is the radius within which the mean cluster overdensity is $\Delta$ times $\rho_{\rm c} (z)$. The relation with the concentration $c_{\Delta}$ and the scale radius $r_{\rm s}$ holds by definition of the NFW mass profile \citep{Navarro1997}. In order to change the mass definition to the one of the mean background density one has to scale $\Delta=200$ by $\Omega_{\rm m}(z)=\Omega_{\rm m} (1+z)^3 / H^2(z)$.
Due to the fact that $M_{\Delta} / (R_{\Delta}^3 \Delta)$ is constant by definition one can write
\begin{equation}
\frac{M_{\Delta}}{c_{\Delta}^3} = \frac{M_{200}}{c_{200}^3} \frac{\Delta}{200}, 
\end{equation}
where $c_{\Delta}$ and $c_{200}$ are related through the assumed NFW mass density profile by
\begin{equation}
\left( \frac{c_{200}}{c_{\Delta}} \right)^3 
\frac{\ln (1+c_{\Delta}) - c_{\Delta}/(1+c_{\Delta}) }
{\ln (1+c_{200})-c_{200}/(1+c_{200})} = \frac{\Delta}{200}.
\label{eqa:rdelta}
\end{equation}
Together with the $c$-$M$ relation of \cite{Zhao2009}, which we adopted for this work, one can directly scale from $M_{200\rm c}$ to $M_{200\rm m}$. For the selected clusters the results are given in the third column of Tab.~\ref{tab:clusters}. \\

Since the mass function is very steep at the high mass end, it is more likely that lower mass systems scatter up than higher mass systems scatter down, resulting in a systematic shift. This effect is known as Eddington bias \citep{Eddington1913} and has to be corrected for when observed masses have to be related to the distribution functions of the most massive halo. This is done, following \cite{Mortonson2011}, by shifting the observed mass, $M_{\rm obs}$, to a corrected mass, $M_{\rm corr}$, by
\begin{equation}
\ln M_{\rm corr}=\ln M_{\rm obs}+\frac{1}{2}\epsilon\sigma_{\ln M}^2,
\end{equation} 
where $\epsilon$ is the local slope of the mass function ($\dd n/\dd \ln M\propto M^\epsilon$) and $\sigma_{\ln M}$ is the uncertainty in the mass measurement. For the systems listed above we inferred $\sigma_{\ln M}$ from the reported observational limits on the mass and rounded them up, just to be on the conservative side. The adopted values for $\sigma_{\ln M}$ are listed together with the Eddington bias corrected masses in Tab.~\ref{tab:clusters} for all eight studied clusters. The variations in $\sigma_{\ln M}$ stem from the fact that some masses were estimated by the combination of different probes, like Lensing+SZ+X-ray, whereas some others are based only on a single probe.  \\
The two corrections, the change of mass definition and the Eddington bias, work in opposite directions, because the former increases the cluster mass and the latter lowers it. The impact of the different mass definition is stronger at low redshifts, since at higher redshift $\Omega_{\rm m}(z)\sim 1$, resulting in a correction of $5-6$ per cent for the high-$z$ clusters in comparison to $\sim 36$ per cent for A2163. For a statistical analysis it is crucial to take both effects into account or the drawn conclusions might be wrong. 
\subsection{Application to single clusters}\label{subsec:high_z}
After having introduced all the effects that have to be taken into account for a sound statistical analysis of the existence probability of single very massive clusters, we now apply GEV to the four most massive clusters in the four redshift intervals summarised in Tab.~\ref{tab:bins}. In order to demonstrate the impact of the bias discussed in Section~\ref{subsec:bias}, we computed the GEV distributions for both choices, a priori and a posteriori, of the redshift intervals. The results of these calculations are shown for the four clusters A2163, ACT-CL~J0102, SPT-CL~J2106 and XXMU~J0044 in Fig.~\ref{fig:spt} from top to bottom. For each cluster the calculations are based on the corresponding survey area ranging from $A_s=27\,490\,{\rm deg}^2$ for A2163 to $A_s=80\,{\rm deg}^2$ for XXMU~J0044. The survey areas are also given in each left-hand panel of Fig.~\ref{fig:spt}. In the figures, the red, dash-dotted curves show the distributions for the biased case with $z\in[z_{\rm obs},3.0]$ and the black, solid lines for the a priori defined redshift bins, as denoted in the key. It can be nicely seen in Fig.~\ref{fig:spt} how, in the biased case, the clusters move towards the tail of the distribution and appear therefore less likely to be found. Depending on at what redshift the cluster of interest resides at in the redshift bin, the effect of the bias is more or less pronounced, as can be seen by comparing the PDF of A2163 with the ones of ACT-CL~J0102 or SPT-CL~J2106. The bias is substantial in the sense that the existence probabilities increase and the allowed range due to the mass error widens, as is illustrated by the red and grey areas shown in the right-hand panels of Fig.~\ref{fig:spt} for the CDFs. The reason for the big difference in the allowed probability region is the steepness of the CDF of the most massive cluster in the given survey volume.\\
For the probability to find the most massive cluster in a given bin to be more massive than the observed one, we obtain the following results: $Pr\lbrace m>M_{\rm obs}\rbrace = 57.2$ per cent for A2163 in bin $1$, $Pr\lbrace m>M_{\rm obs}\rbrace = 38.1$ per cent for ACT-CL~J0102 in bin $2$, $Pr\lbrace m>M_{\rm obs}\rbrace = 22.6$ per cent for SPT-CL~J2106 in bin $3$ and $Pr\lbrace m>M_{\rm obs}\rbrace  = 14.4$ per cent for XXMU~J0044 in bin $4$. The allowed range in the existence probability due to the uncertainty in the cluster mass determination is listed in the rightmost column of Tab.~\ref{tab:bins}.\\
This result clearly states that none of the very massive clusters reported so far can be considered to be in tension with $\Lambda$CDM, even when the upper allowed mass limit is taken. At this point, our findings differ significantly from the results presented in \cite{Chongchitnan2011}\footnote{The authors include also the effect of the halo  bias in their calculations. However, \cite{Davis2011} showed that the difference between the asymptotic form introduced in Section~\ref{sec:GEV} and the full calculation is not very large on large scales (the position of the peak and the tail behaviour match rather well). Furthermore, this difference is certainly smaller than needed to explain the strong discrepancy with respect to our results.}, who find that XXMU~J0044 resides in the extreme tail of the GEV distribution, even when taking the full-sky as survey area. This is in contradiction with our results according to which, even for $A_s=80\,{\rm deg}^2$, we find a probability of existence of $~14.4$ per cent. In our study we neglected any overlap of the XDCP survey footprint with other surveys in the same area of the sky: taking this into account would increase the effective survey area and thus also the probability of existence. An alternative argument is based on a simple calculation of the number of clusters $N(>m)$ more massive than $m$ for the reported mass range and a redshift of $z\geq 1.579$. The results range from $\sim 7-80$ (or even $30-250$, when using the \cite{Sheth&Tormen1999} mass function as done by the authors), supporting our findings that this system can be considered as perfectly normal and expected in a $\Lambda$CDM Universe. In view of this, the claim that this system is an indicator for non-Gaussianity on the level of $f_{\rm NL}\simeq 360$ seems to be highly questionable\footnote{We would also like to point out that a likelihood analysis based on a single data point and the inference of the most likely $f_{\rm NL}$ from it can be highly misleading.}.  \\
%--------------------------------------------
\begin{table}
\centering
\caption{Selection of the most massive clusters in the four redshift intervals and the inferred probability of existence.}
\begin{tabular}{lcccc} \hline
bin & $z$-intervall & cluster & $Pr\lbrace m>M_{\rm obs}\rbrace$ & range  \\
\hline
$1$ & $0\le z\le 0.5$ &  A2163  &  $57.2\%$ & $(11.3-99)\%$\\
$2$ & $0.5\le z\le 1.0$ & ACT-CL~J0102  & $38.1\%$  & $(9.9-83.2)\%$\\
$3$ & $1.0\le z\le 1.5$  & SPT-CL~J2106 &  $22.6\%$ & $(5.2-66.7)\%$\\
$4$ & $1.5\le z\le 3.0$ & XXMU~J0044  &  $14.4\%$ & $(4.2-39.9)\%$\\
\hline
\end{tabular}\label{tab:bins}
\end{table}
%--------------------------------------------
%-------------------------------------
\begin{figure}
\centering
\includegraphics[width=0.9\linewidth]{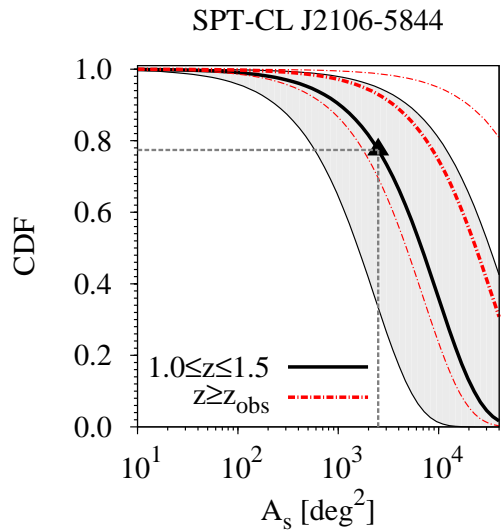}
\caption{CDF as a function of survey area, $A_{\rm s}$, with fixed observed masses, $M_{200\rm m}^{\rm Edd}$, for SPT-CL~J2106. The grey shaded areas show the uncertainty in the mass measurements and the black triangle denotes the \textit{SPT} survey area of $A_{\rm s} =  2\,500\,{\rm deg}^2$. The red, dash-dotted line shows the biased results based on a posteriori choice of the $z$-interval and the black, solid lines are for an a priori fixed interval of $z\in [1.0,1.5]$.}\label{fig:new}
\end{figure}
%-------------------------------------
For the GEV analysis performed in this section, we compared the most massive observed cluster with the theoretical distribution function. In doing so, it is assumed that the observed most massive cluster is also the true most massive cluster in the volume of interest. In the lowest redshift bin of $0\le z\le 0.5$ for instance, all four low-$z$ clusters from Tab.~\ref{tab:clusters} exhibit potential overlap in their allowed masses, such that e.g. the true most massive cluster in the redshift interval could be A370 instead of A2163. However, since all the clusters in this redshift interval are very likely to be found in $\Lambda$CDM, the results would remain unchanged even if another cluster than A2163 was the true most massive one. For the high redshift intervals, the chosen systems are so extreme in their respective survey areas that it is very unlikely that another system is the true most massive one. In general, one should keep in mind that the probabilities stated above are grounded on the assumption that the individual cluster is the true most massive one in the volume. If this is not the case the exact numbers from above would change, but the conclusion that none of them is in significant tension with $\Lambda$CDM would still hold. The same statement is also valid for a different a priori choice of the redshift intervals.
%---------------------------------------------------------------
\section{Comparing high-$z$ with low-$z$ clusters}\label{sec:lowVShigh}
%---------------------------------------------------------------
%-------------------------------------
\begin{figure*}
\centering
\includegraphics[width=0.245\linewidth]{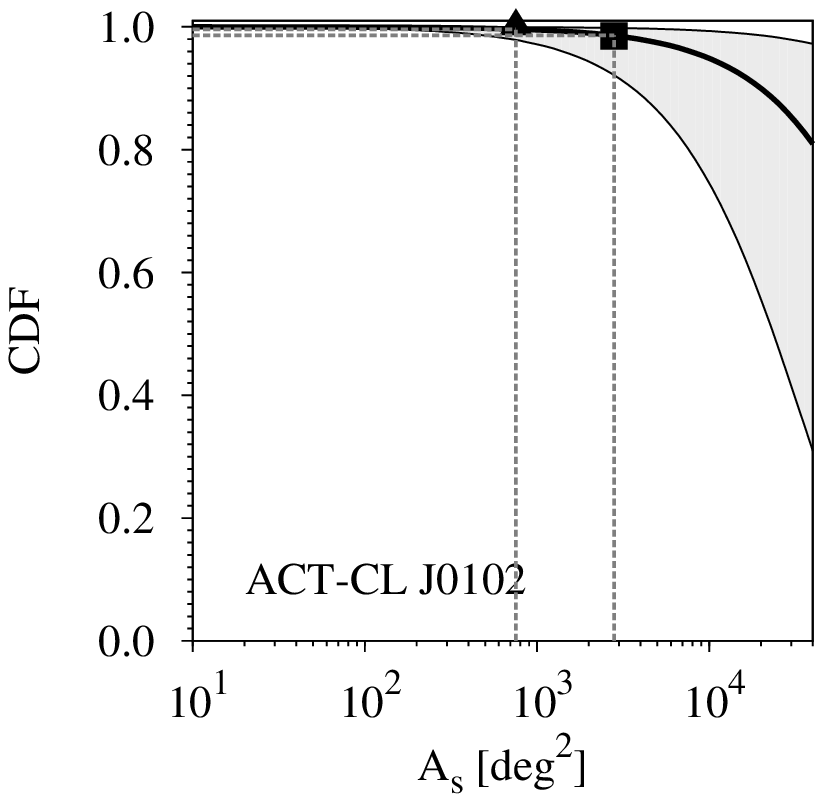}
\includegraphics[width=0.245\linewidth]{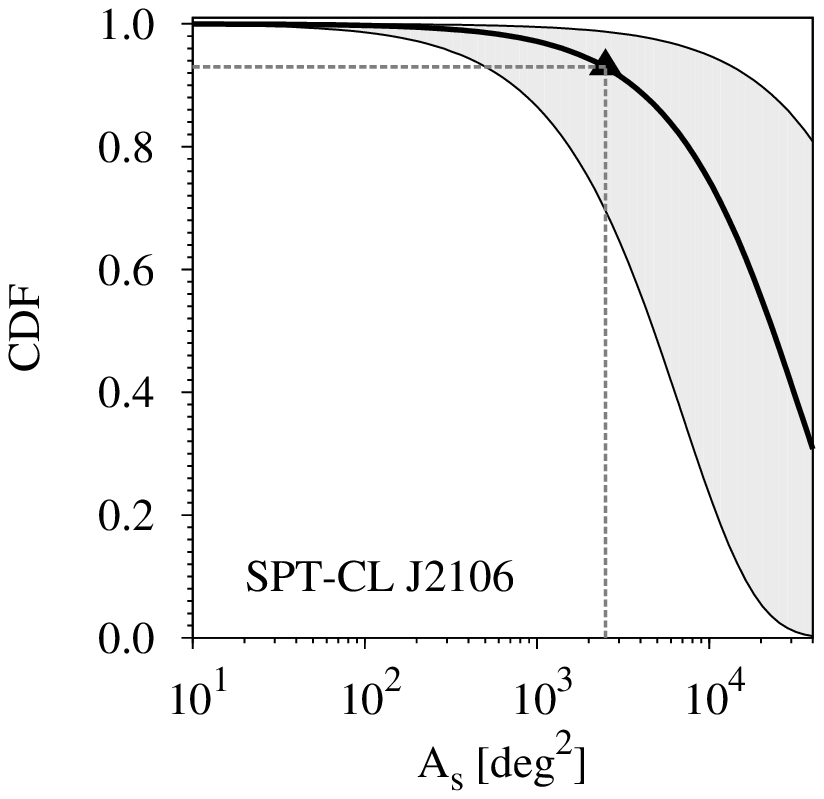}
\includegraphics[width=0.245\linewidth]{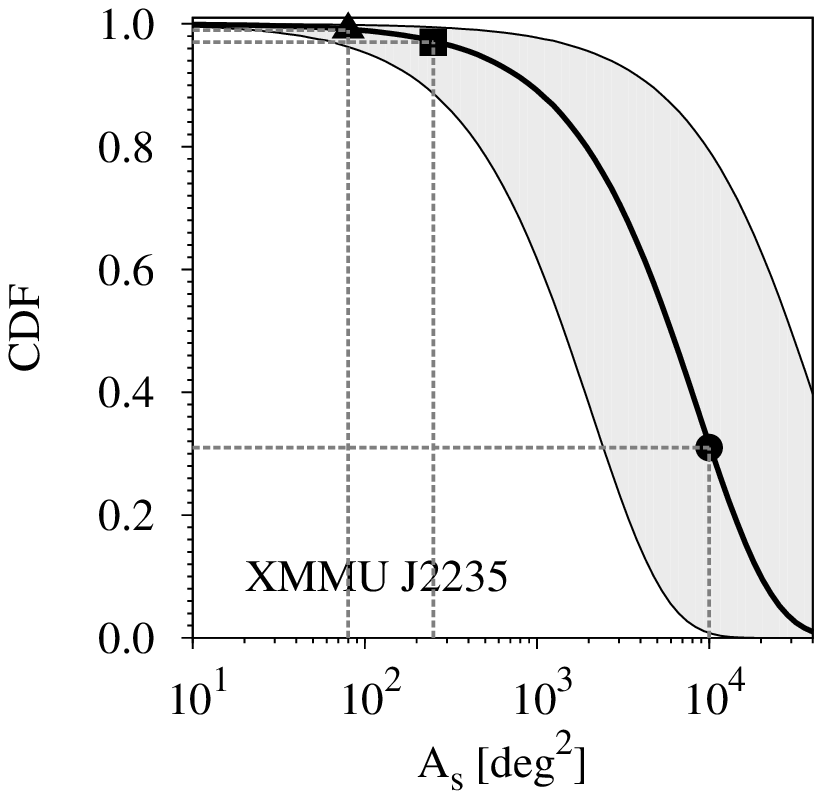}
\includegraphics[width=0.245\linewidth]{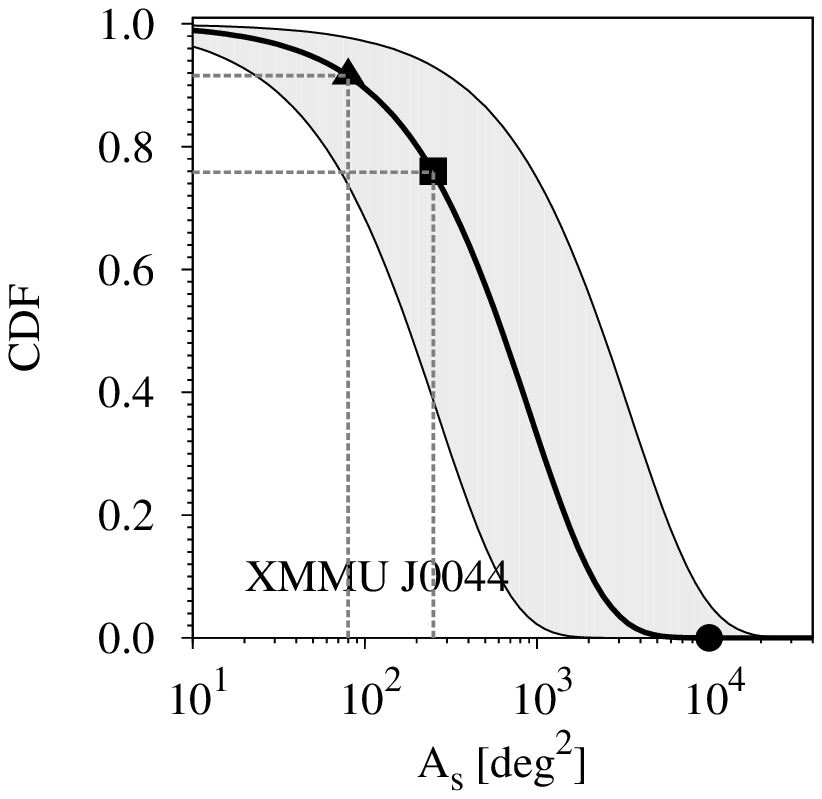}\\
\includegraphics[width=0.245\linewidth]{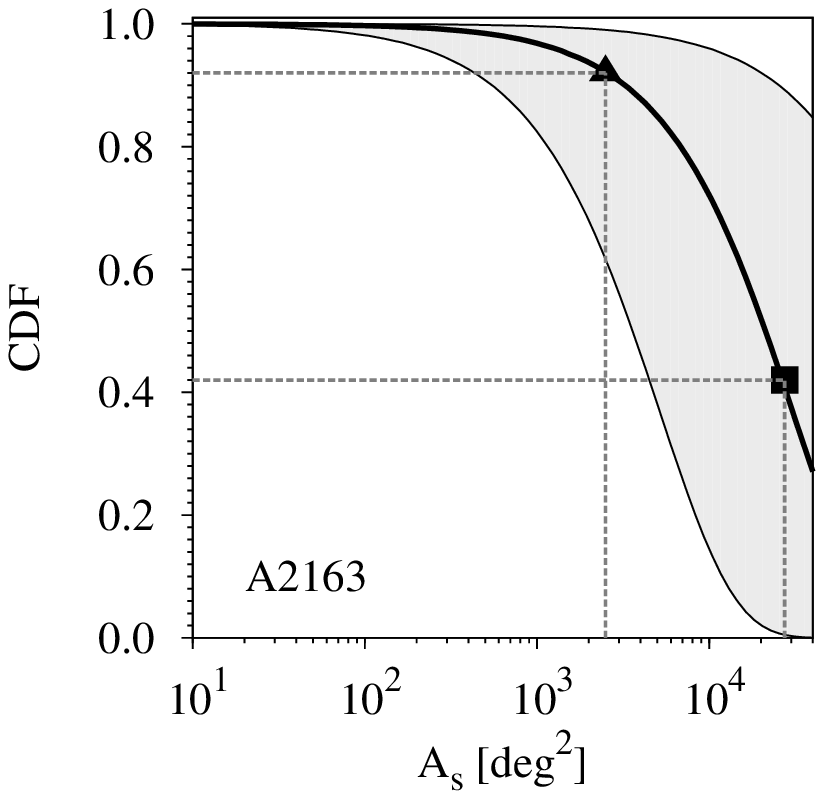}
\includegraphics[width=0.245\linewidth]{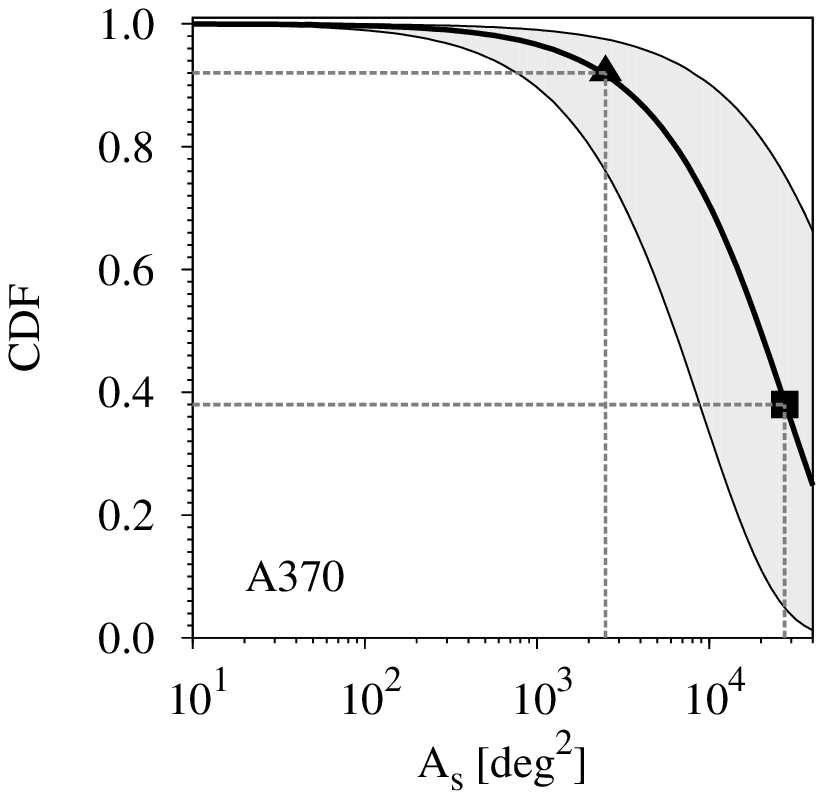}
\includegraphics[width=0.245\linewidth]{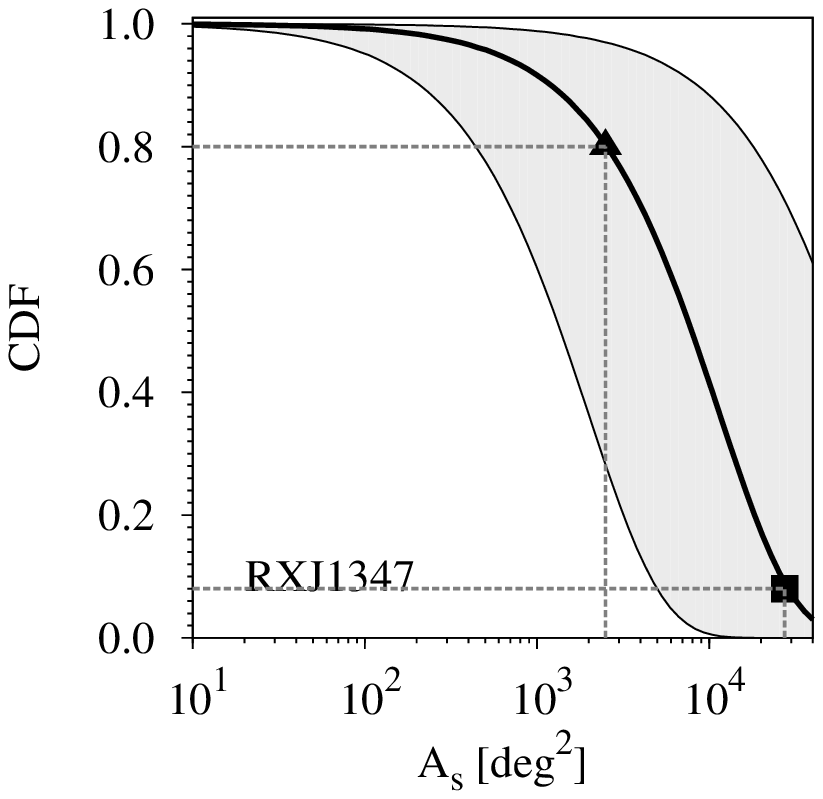}
\includegraphics[width=0.245\linewidth]{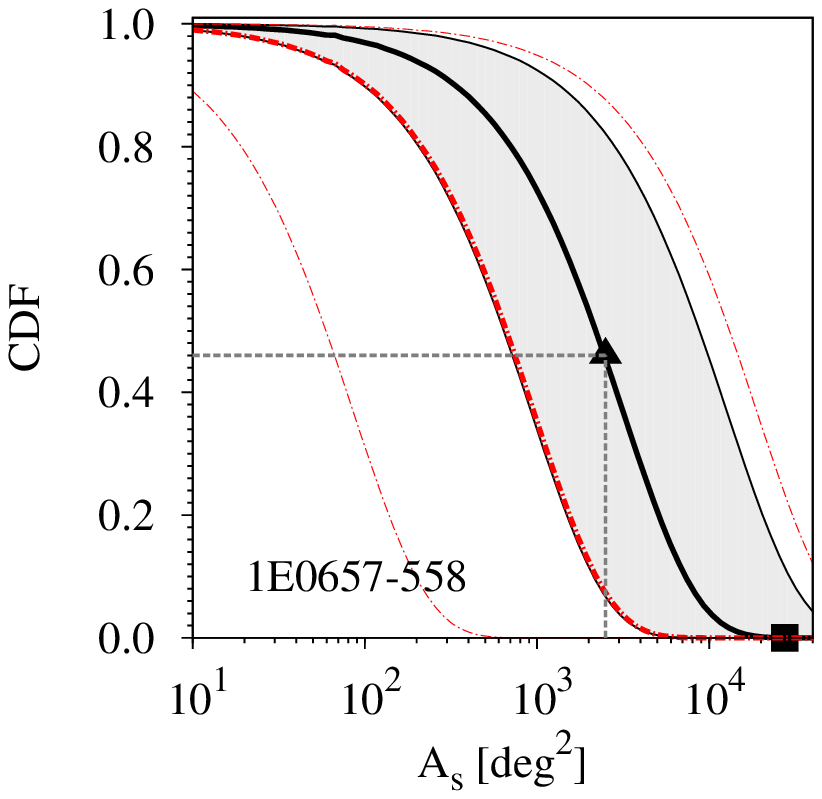}
\caption{Biased CDF as a function of survey area, $A_{\rm s}$, with fixed observed masses, $M_{200\rm m}^{\rm Edd}$, for all eight high-$z$ (upper row) and low-$z$ (lower row) clusters arranged from left to right and top to bottom: ACT-CL~J0102, SPT-CL~J2106, XMMU~J2235, XMMU~J0044, A2163, A370, RXJ1347 and 1E0657. The grey shaded areas show the uncertainty in the mass measurements and the different symbols denote particular choices of the survey area which are: $A_{\rm s} = 775$ (triangle) and $2\,800\,{\rm deg}^2$ (square) for ACT-CL~J0102; $A_{\rm s} = 2\,500\,{\rm deg}^2$ (triangle) for SPT-CL~J2106; $A_{\rm s} = 80$ (triangle) , $250$ (square) and $10\,000\,{\rm deg}^2$ (circle) for XMMU~J2235 and XMMU~J0044. And $A_{\rm s} = 2\,500$ (triangle) and $27490\,{\rm deg}^2$ (square) for all four low-$z$ systems in the lower row. For 1E0657, the red, dash-dotted curves in in the lower left-hand panel show the results based on the SZ mass \citep{Williamson2011} and the black, solid lines the results for the X-ray mass \citep{Maughan2011}.}\label{fig:test}
\end{figure*}
%-------------------------------------
The existence probabilities listed in Tab.~\ref{tab:bins} seem to decrease with increasing redshift such that high-$z$ clusters appear to be rarer than low-$z$ ones. This could naively be interpreted as a signature of a modification of the growth of structure, or as a substantial bias of the high-$z$ cluster mass estimates. In the following we will address this point in more detail.
%---------------------------------------------------------------
\subsection{Impact of ill-defined survey areas}
%---------------------------------------------------------------
For a statistical analysis of high-$z$ clusters, the exact definition of the survey area is crucial if conclusions regarding the cosmological model have to be drawn. For a given mass of a cluster, a too small area will lead to a shift of the most likely maximum mass to smaller masses and, hence, amplify a possible discrepancy with respect to the fiducial $\Lambda$CDM model. A too large survey area would shift the peak of the PDF to the right and therefore alleviate possible tension with $\Lambda$CDM. Particularly at high redshifts, it can be difficult to define the proper survey area since the cluster of interest might have been observed by multiple surveys with different selection functions and observing strategies.\\
Instead of plotting the CDFs and PDFs as a function of mass and comparing an observed cluster with it, as presented in Fig.~\ref{fig:spt}, it is more meaningful to present the CDF for a fixed observed mass as a function of survey area. This is shown in Fig.~\ref{fig:new} for SPT-CL~J2106 for both the biased (red, dash-dotted lines) and the unbiased (black, solid lines) case. The grey shaded areas represent the range in the CDF allowed by the uncertainty in the mass estimates. It can directly be seen how the cluster becomes more likely to be found as the survey area increases. As long as the survey areas are not clearly defined, the position of the curve is the best indicator of how extreme a cluster is. The further right (larger $A_{\rm s}$) the curve drops to zero, the more extreme the cluster is.\\
For instance, from Fig.~\ref{fig:new} we can infer that if SPT-CL~J2106 would have been detected in a fictive survey area of $100\,{\rm deg}^2$, the assigned probability to find this cluster would be less than $1$ per cent. However, detected in the \textit{SPT} field of $A_{\rm s} =  2\,500\,{\rm deg}^2$, the probability increases already to $22.6$ per cent and it would increase further if the survey area had been increased without finding a more massive object in the redshift range. \\
In order to compare our GEV based analysis with the results of \cite{Foley2011}, who did not take the bias discussed in \cite{Hotchkiss2011} into account, one has to look at the red, dash-dotted line in Fig.~\ref{fig:new}. In doing so and by taking the exact reported mass of SPT-CL~J2106, we obtain a biased probability of $\lesssim 7$ per cent to find such a system in the erroneously a posteriori fixed redshift interval. It agrees well with the findings of \cite{Foley2011}, who report a probability of $\lesssim 5$ per cent for such a system based on a full likelihood calculation utilising a MCMC method \citep{Vanderlinde2010, Lewis2002} for the \textit{WMAP}7 cosmology. We obtain a slightly higher probability due to our conservative choice of $\sigma_{\ln M}$, resulting in a larger correction for the Eddington bias towards smaller masses. This result shows that GEV could serve as a tool for quick analysis of extreme clusters at high redshifts before a full likelihood analysis is performed, particularly from the point of view that it can easily be adapted to alternative cosmological models, like e.g. models with non-Gaussianity \citep{Chongchitnan2011} or quintessence models \citep{Waizmann2011}.
%---------------------------------------------------------------
\subsection{Comparison of all eight clusters}
%---------------------------------------------------------------
In the previous section, we showed that the apparent decrease of the existence probability with increasing redshift could very likely be explained by the relatively small survey areas at high redshifts. In order to study this in more detail, we decided to perform a GEV analysis for all eight clusters discussed in Section~\ref{sec:GEVclust}. However, we do not correct for the bias discussed in Section~\ref{subsec:bias} and choose the redshift intervals a posteriori to be $z\in [z_{\rm obs},3.0]$. The results of this calculation are shown in Fig.~\ref{fig:test}, again in the form of the CDF as function of the survey area, for the four high-$z$ clusters in the upper row and for the four low-$z$ ones in the lower row. It is important to note that now the CDF can no longer be interpreted as a measure of the unbiased probability of tension with $\Lambda$CDM, but only as a measure of relative extremeness.\\
Even with the biased measure, none of the clusters is particularly in tension with $\Lambda$CDM, but they can all be considered as rather perfectly normal and expected systems. When confronting the low-$z$ with the high-$z$ systems on the basis of the CDF as a function of the survey area, the relative positions of the curves do not give any support for the claim that clusters at high-$z$ or more extreme than low-$z$ ones.\\
If A2163, A370 and RXJ1347 had been observed in a \textit{SPT}-like field (triangle) instead of the \text{RASS} one (square), the resulting biased probabilities of their existence would be similar to the high-$z$ ones (especially when taking into consideration that larger $\sigma_{\ln M}$ values lead to smaller Eddington bias corrected masses for the low-$z$ systems). And, vice versa, an increase of the survey area for the high-$z$ systems would alleviate their low probabilities of existence, unless more extreme objects are found in the future.\\
By looking to currently known very massive single objects at low and high redshifts and analysing their CDFs as function of area, one does not find any evidence for trend with redshift that high-$z$ systems are more extreme than low-$z$ systems. Therefore there is no indication for deviations from the $\Lambda$CDM expectations so far. Of course, when drawing conclusions from this fact, one has to be careful, since at high-$z$ the sufficiently deep covered survey area is only a fraction of the low redshift one. Furthermore, cluster mass estimates, particularly at high redshifts, are delicate and possible biases at high-$z$ are not yet fully explored. As a verdict it is certainly interesting to confront the most massive clusters at different redshifts with the expectations from theory once a sufficiently deep full-sky cluster survey will be available. 
%-------------------------------------
\begin{figure*}
\centering
\includegraphics[width=0.33\linewidth]{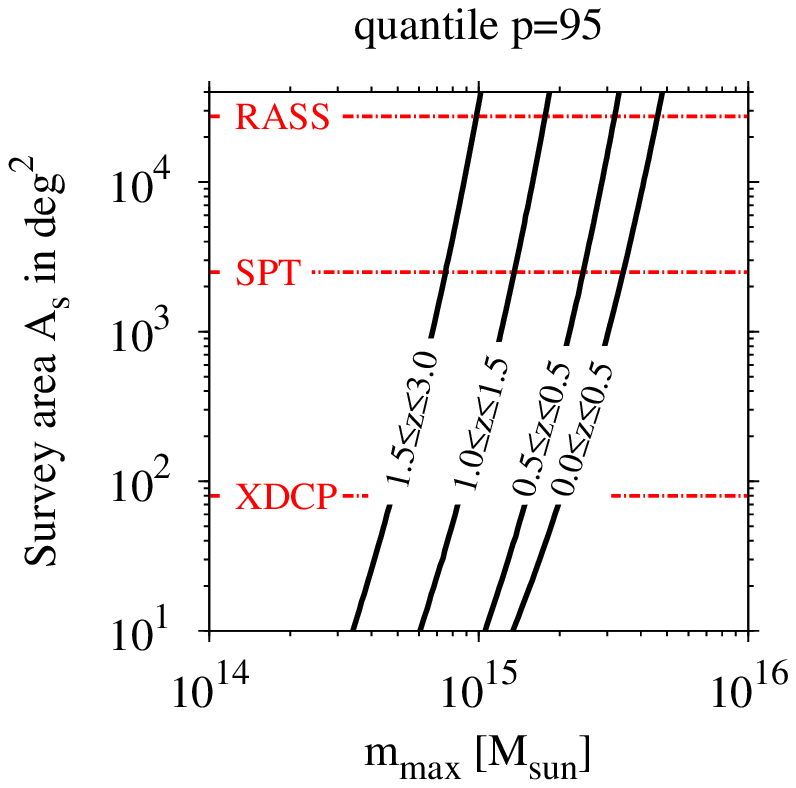}
\includegraphics[width=0.33\linewidth]{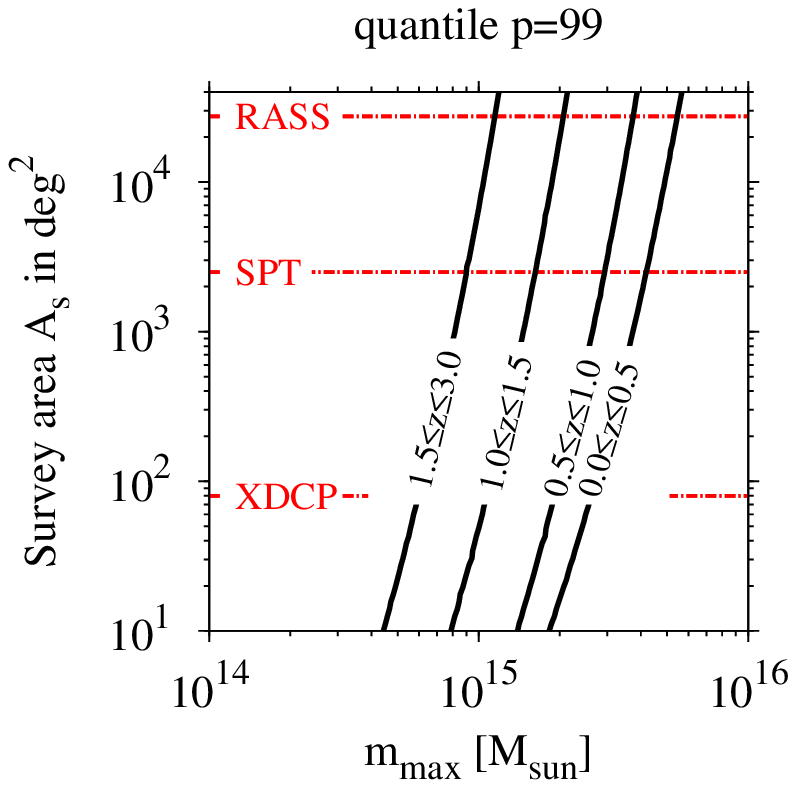}
\includegraphics[width=0.33\linewidth]{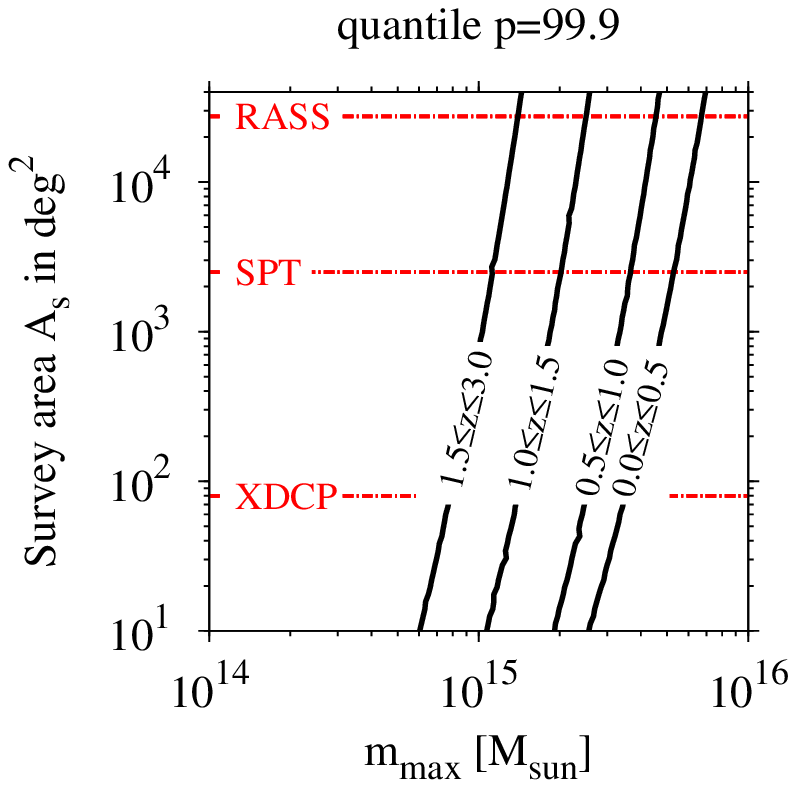}
\caption{Contour plots for a fixed $p=95,\,99$ and $99.9$ (from left to right) in the plane spanned by the cluster mass and survey area for four different redshift intervals as denoted on each iso-contour. For better comparison the survey areas of the \textit{RASS}, \textit{SPT} and \textit{XDCP} are illustrated by the red, dash-dotted lines.}\label{fig:contours}
\end{figure*}
%------------------------------------
%------------------------------------------------------
\section{Extreme quantiles}\label{sec:quantiles}
%------------------------------------------------------
All the probability ranges stated in the previous section are very closely related to the statistical concept of quantiles. The extreme quantile function based on GEV is the inverse of the CDF from equation \eqref{eq:GEV} and is found to be given by
\begin{equation}
x_p = \left\{ 
  \begin{array}{l l}
     \alpha-\frac{\beta}{\gamma}\lbrace 1-\left[-\ln\left(\frac{p}{100}\right)\right]^{-\gamma}\rbrace, & \quad {\rm for}\quad\gamma\neq 0,\\
    \alpha-\beta\ln\left[-\ln\left(\frac{p}{100}\right)\right], & \quad {\rm for}\quad\gamma = 0,\\
  \end{array} \right.
\end{equation}
A quantile $x_p$ is the value of the random variable for which the probability $Pr\lbrace X_i\le x_p\rbrace$ is $p$ or, in our case of interest, it is the mass of a cluster such that, in $p$ percent of the observed patches, the most massive cluster has a smaller mass than the $p$-quantile mass. Usually $x_p$ is also referred to as \textit{return level} and it is connected to the \textit{return period} $\tau_p=1/(1-p/100)$.\\
For the case studied here, it is interesting to compute the lines for the four a priori defined redshift intervals (see Tab.~\ref{tab:bins}) in the plane spanned by mass and survey area that would correspond to a given fixed quantile $p$. The results are shown in Fig.~\ref{fig:contours} for $p=95$ (left-hand panel), $p=99$ (central panel) and for $p=99.9$ (right-hand panel), where the red, dash-dotted lines denote from top to bottom the survey areas of the \textit{RASS}, \textit{SPT} and \textit{XDCP} surveys. It can be seen that the contours for fixed redshift bins and quantile $p$ exhibit a steep increase with mass and they also reflect the fact that, in high-$z$ bins, smaller masses are needed to reach a given quantile.\\

The contour plots in Fig.~\ref{fig:contours} confirm the results from Section~\ref{subsec:high_z} in which we found that none of the very massive known clusters, neither at high nor at low redshifts, is in tension with $\Lambda$CDM. If we define the $p=99.9$ quantile as the line where the $\Lambda$CDM model is excluded and we assume a \textit{RASS}-like cluster survey, then one would need to find a cluster of $M_{200\rm m}>1.4\times 10^{15}\,M_\odot$ at $z>1.5$. For the third redshift bin of $1.0\le z\le 1.5$, one would require a mass of $M_{200\rm m}>2.5\times 10^{15}\,M_\odot$, which is roughly the double of the mass of SPT-CL~J2106. For the lowest redshift bins of $0.5\le z\le 1.0$ and $0.0\le z\le 0.5$, we find $M_{200\rm m}>4.5\times 10^{15}\,M_\odot$ and $M_{200\rm m}>6.7\times 10^{15}\,M_\odot$, respectively. The existence of the latter can already nowadays be excluded on the basis of \textit{RASS} and \textit{MACS} \citep{Ebeling2001} surveys and for redshifts below $z=1$ \textit{PLANCK} \citep{Tauber2010} should be capable of delivering a definite answer. For the high redshift end with $z>1$, future missions like \textit{EUCLID} \citep{Laureijs2011} and \textit{eROSITA} \citep{Cappelluti2011} should definitely be able to detect such extreme systems if they exist.\\

The lesson to be learnt from the exercise above is that GEV could offer a quick way for characterising the statistical expectations for very massive high-$z$ clusters. But for cosmological model testing, a proper knowledge of the survey area is crucial. Of course, it is not only the area but also the depth that matters because other surveys covering the same object should ideally be as deep as the original survey to be matchable, which usually is a particular problem for the very high-$z$ clusters. Furthermore, even if one finds a high quantile cluster, let's assume $p>99$ in a small survey area, one should keep in mind that, when sampling a GEV distribution, such systems have to exist if the sample is big enough. Considering the current high-$z$ coverage this is all that can currently be done, but with future full-sky surveys it might be interesting to try to measure the CDF itself as proposed in \cite{Waizmann2011}.
%------------------------------------------------------
\section{Summary and conclusions}\label{sec:conclusions}
%------------------------------------------------------
In this work we presented the application of general extreme value statistics on very massive single clusters at high and low redshifts. After introducing the formalism, we applied these statistics to eight very massive clusters at high and low redshifts. On the high redshift side, those clusters comprise the \textit{ACT} detected cluster ACT-CL~J0102-4915, the \textit{SPT} cluster SPT-CL~J2106-5844 and two clusters found by the \textit{XMM-Newton} Distant Cluster Project (\textit{XDCP}) survey: XMMU~J2235.32557 and XMMU~J0044.0-2033. For the low redshift systems, we considered  A2163, A370, RXJ1347-1145 and 1E0657 where the latter was added to the study because it is the most massive cluster in the \textit{SPT} survey.\\
By computing the CDFs and PDFs for the individual survey set-ups as well as fictitious survey areas, we relate the individual systems with the probability to find them as most massive system for a given size of the survey area. In order to avoid the bias arising from the posterior choice of the redshift interval, as discussed in \cite{Hotchkiss2011}, we define the redshift intervals prior to the analysis to be: $0.0\le z\le 0.5$, $0.5\le z\le 1.0$, $1.0\le z\le 1.5$ and $1.5\le z\le 3.0$. From the aforementioned eight clusters, we chose the respective most massive one in each individual bin and calculate its existence probability in a $\Lambda$CDM cosmology.\\
We find that, in the lowest redshift bin of $0.0\le z\le 0.5$ and the \textit{RASS} survey area of  $27\,490\,{\rm deg}^2$, A2163 has an existence probability of $\sim 57$ per cent in a range of $\sim(11-99)$ per cent reflecting the uncertainty in the mass determination. In the second redshift interval of $0.5\le z\le 1.0$,  ACT-CL~J0102 has, in the combined survey area of \textit{ACT} and \textit{SPT} of $2\,800\,{\rm deg}^2$, an existence probability of $\sim 38$ per cent in a range of $\sim(10-83)$ per cent. For SPT-CL~J2106-5844, we find $\sim 23$ per cent in a range of $\sim(5-67)$ per cent in the interval of $1.0\le z\le 1.5$ and the \textit{SPT} survey area of $2\,500\,{\rm deg}^2$. In the highest redshift bin of $1.5\le z\le 3.0$, we find for XMMU~J0044.0-2033 in the $80\,{\rm deg}^2$ \textit{XDCP} survey area a probability of $\sim 14$ per cent in a range of $\sim(4-40)$ per cent. This result, by classifying this system as not in tension with $\Lambda$CDM at all, differs significantly from the results of \cite{Chongchitnan2011} who claim this system to be in extreme tension with $\Lambda$CDM even if the full-sky is assumed as survey area. Therefore, none of the clusters can be considered to be in tension with $\Lambda$CDM even if the upper allowed observed mass is assumed. \\
If we neglect the aforementioned bias arising from the posterior choice of the redshift interval, as it has been done in most of the literature in the past years (apart from \cite{Hotchkiss2011} and \cite{Hoyle2011b}), and fix the redshift interval a posteriori, we find for SPT-CL~J2106-5844 a probability of $\lesssim 7$ per cent. This result is in good agreement with the findings of \cite{Foley2011}, who report a probability of $\lesssim 5$ per cent based on a full likelihood analysis. The agreement of the two results advocates GEV as a quick tool for analysing the probability of observed high-mass clusters before a full likelihood analysis is performed. \\
By confronting the results of our survey area based GEV analysis for both, low and high redshift clusters, we do not find any indication for a trend in redshift in the sense that high-$z$ systems are more extreme than low-$z$ ones. By studying the CDFs of the observed clusters as a function of the survey area, we show that the most likely explanation for the lower existence probabilities of current high mass, high-$z$ clusters with respect to low-$z$ ones, is the lack of sufficiently deep and large high-$z$ surveys. Thus, based on current data, we do not see any tendency for a deviation from $\Lambda$CDM for the most massive clusters as a function of redshift. Of course, one should also be aware that, apart from the fact that the current observational data are not deep and complete enough, also possible biases can be present in the cluster mass estimates, particularly at high redshifts.\\

In addition, we introduced the extreme quantiles and calculated the contours corresponding to a fixed $p$-quantile for a fixed redshift interval in the plane spanned by cluster mass and survey area. These contours allow one to infer what mass a cluster would need to have in order to cause a substantial tension with $\Lambda$CDM.  We find that a cluster with an existence probability of $1:1000$ would have to have at least a mass of $M_{200\rm m}>4.5\times 10^{15}\,M_\odot$ in $0.5\le z\le 1.0$, $M_{200\rm m}>2.5\times 10^{15}\,M_\odot$ in $1.0\le z\le 1.5$ and $M_{200\rm m}>1.4\times 10^{15}\,M_\odot$ for $z>1.5$. The first two correspond to twice the mass of the most massive currently known cluster in these redshift intervals and the last one corresponds to a cluster similar to SPT-CL~J2106 at $z>1.5$. These very high masses make more and more questionable the possibility to find a single cluster with the potential to significantly question $\Lambda$CDM. However, ongoing and future large area surveys like \textit{PLANCK}, \textit{eROSITA} and \textit{EUCLID} have the capabilities to give a definite answer to this question.\\

Thus, the main conclusions that can be drawn from this work can be summarised as follows:
\begin{enumerate}
\item None of the currently known very massive clusters at high and low redshifts exhibits tension with $\Lambda$CDM.
\item There is no indication for very high-$z$ clusters being more extreme than low-$z$ ones and therefore no indication of deviations from the $\Lambda$CDM structure growth or of strong redshift depended biases in the cluster mass estimates.
\item Clusters with the potential to significantly question $\Lambda$CDM would require substantially higher masses in the different redshift regimes than currently known.
\item GEV is a valuable tool to understand the rareness of massive galaxy clusters and delivers comparable results to more costly full likelihood analyses.  
\end{enumerate}
As a closing word of warning, one should be cautious when trying to reject a distribution function and thus a underlying cosmological model by means of a single observed object. Very rare realisations of a random variable do have to exist and by observing just one there is no telling whether it stems from the tail of the assumed PDF or from the peak of another one. For small patches, one may solve this dilemma by observing many of those and directly reconstruct the underlying CDF \citep{Waizmann2011}; for large patches, however, one is limited by cosmic variance. 

\section*{Acknowledgments}
We would particularly like to thank the referee Shaun Hotchkiss. His valuable comments and remarks helped to substantially improve this manuscript. The authors acknowledge financial contributions from contracts ASI-INAF I/023/05/0, ASI-INAF I/088/06/0, ASI I/016/07/0 COFIS, ASI Euclid-DUNE I/064/08/0, ASI-Uni Bologna-Astronomy Dept. Euclid-NIS I/039/10/0, and PRIN MIUR Dark energy and cosmology with large galaxy surveys. 

\bibliographystyle{mn2e}

\begin{thebibliography}{}

\bibitem[{{Arnaud} {et~al.}(1992){Arnaud}, {Hughes}, {Forman}, {Jones},
  {Lachieze-Rey}, {Yamashita}, \& {Hatsukade}}]{Arnaud1992}
{Arnaud}, M., {Hughes}, J.~P., {Forman}, W., {Jones}, C.,
  {Lachieze-Rey}, M., {Yamashita}, K., {Hatsukade}, Y., 1992, \apj, 390, 345
  
  \bibitem[{{Arnaud} {et~al.}(2005){Arnaud}, {Pointecouteau}, \&
  {Pratt}}]{Arnaud2005}
{Arnaud}, M., {Pointecouteau}, E., {Pratt}, G.~W., 2005, \aap, 441, 893

\bibitem[{{Baldi}(2011)}]{Baldi2011b}
{Baldi}, M., 2011, preprint (arXiv: 1107.5049)

\bibitem[{{Baldi} \& {Pettorino}(2011)}]{Baldi2011a}
{Baldi}, M., {Pettorino}, V., 2011, \mnras, 412, L1

\bibitem[{{Bhavsar} \& {Barrow}(1985)}]{Bhavsar1985}
{Bhavsar}, S.~P., {Barrow}, J.~D., 1985, \mnras, 213, 857

\bibitem[{{Broadhurst} {et~al.}(2008){Broadhurst}, {Umetsu}, {Medezinski},
  {Oguri}, \& {Rephaeli}}]{Broadhurst2008}
{Broadhurst}, T., {Umetsu}, K., {Medezinski}, E., {Oguri}, M., {Rephaeli},
  Y., 2008, \apjl, 685, L9

\bibitem[{{Broadhurst} \& {Barkana}(2008)}]{Broadhurst&Barkana2008}
{Broadhurst}, T.~J., {Barkana}, R., 2008, \mnras, 390, 1647

\bibitem[{{Cappelluti} {et~al.}(2011){Cappelluti}, {Predehl}, {B{\"o}hringer},
  {Brunner}, {Brusa}, {Burwitz}, {Churazov}, {Dennerl}, {Finoguenov},
  {Freyberg}, {Friedrich}, {Hasinger}, {Kenziorra}, {Kreykenbohm}, {Lamer},
  {Meidinger}, {M{\"u}hlegger}, {Pavlinsky}, {Robrade}, {Santangelo},
  {Schmitt}, {Schwope}, {Steinmitz}, {Str{\"u}der}, {Sunyaev}, \&
  {Tenzer}}]{Cappelluti2011}
{Cappelluti}, N., {Predehl}, P., {B{\"o}hringer}, H., {et~al.} 2011, Memorie
  della Societa Astronomica Italiana Supplementi, 17, 159

\bibitem[{{Carlstrom} {et~al.}(2011){Carlstrom}, {Ade}, {Aird}, {Benson},
  {Bleem}, {Busetti}, {Chang}, {Chauvin}, {Cho}, {Crawford}, {Crites}, {Dobbs},
  {Halverson}, {Heimsath}, {Holzapfel}, {Hrubes}, {Joy}, {Keisler}, {Lanting},
  {Lee}, {Leitch}, {Leong}, {Lu}, {Lueker}, {Luong-van}, {McMahon}, {Mehl},
  {Meyer}, {Mohr}, {Montroy}, {Padin}, {Plagge}, {Pryke}, {Ruhl}, {Schaffer},
  {Schwan}, {Shirokoff}, {Spieler}, {Staniszewski}, {Stark}, {Tucker},
  {Vanderlinde}, {Vieira}, \& {Williamson}}]{Carlstrom2011}
{Carlstrom}, J.~E.~{et~al.}, 2011, \pasp, 123,
  568

\bibitem[{{Cay{\'o}n} {et~al.}(2011){Cay{\'o}n}, {Gordon}, \&
  {Silk}}]{Cayon2011}
{Cay{\'o}n}, L., {Gordon}, C., {Silk}, J., 2011, \mnras, 415, 849

\bibitem[{{Chongchitnan} \& {Silk}(2011)}]{Chongchitnan2011}
{Chongchitnan}, S., {Silk}, J., 2011, preprint (arXiv: 1107.5617)

\bibitem[{{Clowe} {et~al.}(2006){Clowe}, {Brada{\v c}}, {Gonzalez},
  {Markevitch}, {Randall}, {Jones}, \& {Zaritsky}}]{Clowe2006}
{Clowe}, D., {Brada{\v c}}, M., {Gonzalez}, A.~H., {Markevitch}, M., {Randall}, S.~W., {Jones}, C., {Zaritsky}, D., 2006, \apjl, 648,
  L109
  
\bibitem[{{Coles}(1988)}]{Coles1988}
{Coles}, P., 1988, \mnras, 231, 125

\bibitem[{{Coles}(2001)}]{Coles2001}
{Coles}, S., 2001, {An Introduction to Statistical Modeling of Extreme Values}
  (Springer)

\bibitem[{{Colombi} {et~al.}(2011){Colombi}, {Davis}, {Devriendt}, {Prunet}, \&
  {Silk}}]{Colombi2011}
{Colombi}, S., {Davis}, O., {Devriendt}, J., {Prunet}, S., {Silk}, J., 2011,
  \mnras, 414, 2436

\bibitem[{{Davis} {et~al.}(2011){Davis}, {Devriendt}, {Colombi}, {Silk}, \&
  {Pichon}}]{Davis2011}
{Davis}, O., {Devriendt}, J., {Colombi}, S., {Silk}, J., {Pichon}, C., 2011,
  \mnras, 413, 2087
  
\bibitem[{{Ebeling} {et~al.}(2001){Ebeling}, {Edge}, \& {Henry}}]{Ebeling2001}
{Ebeling}, H., {Edge}, A.~C., {Henry}, J.~P. 2001, \apj, 553, 668   

\bibitem[{{Eddington}(1913)}]{Eddington1913}
{Eddington}, A.~S., 1913, \mnras, 73, 359

\bibitem[{{Enqvist} {et~al.}(2011){Enqvist}, {Hotchkiss}, \&
  {Taanila}}]{Enqvist2011}
{Enqvist}, K., {Hotchkiss}, S., {Taanila}, O., 2011, \jcap, 4, 17

\bibitem[{{Ettori} {et~al.}(2001){Ettori}, {Allen}, \& {Fabian}}]{Ettori2001}
{Ettori}, S., {Allen}, S.~W., {Fabian}, A.~C., 2001, \mnras, 322, 187

\bibitem[{{Fisher} \& {Tippett}(1928)}]{Fisher1928}
{Fisher}, R., {Tippett}, L., 1928, Proc. Cambridge Phil. Soc., 24, 180

\bibitem[{{Foley} {et~al.}(2011){Foley}, {Andersson}, {Bazin}, {de Haan},
  {Ruel}, {Ade}, {Aird}, {Armstrong}, {Ashby}, {Bautz}, {Benson}, {Bleem},
  {Bonamente}, {Brodwin}, {Carlstrom}, {Chang}, {Clocchiatti}, {Crawford},
  {Crites}, {Desai}, {Dobbs}, {Dudley}, {Fazio}, {Forman}, {Garmire}, {George},
  {Gladders}, {Gonzalez}, {Halverson}, {High}, {Holder}, {Holzapfel}, {Hoover},
  {Hrubes}, {Jones}, {Joy}, {Keisler}, {Knox}, {Lee}, {Leitch}, {Lueker},
  {Luong-Van}, {Marrone}, {McMahon}, {Mehl}, {Meyer}, {Mohr}, {Montroy},
  {Murray}, {Padin}, {Plagge}, {Pryke}, {Reichardt}, {Rest}, {Ruhl},
  {Saliwanchik}, {Saro}, {Schaffer}, {Shaw}, {Shirokoff}, {Song}, {Spieler},
  {Stalder}, {Stanford}, {Staniszewski}, {Stark}, {Story}, {Stubbs},
  {Vanderlinde}, {Vieira}, {Vikhlinin}, {Williamson}, \& {Zenteno}}]{Foley2011}
{Foley}, R.~J.~{et~al.}, 2011, \apj, 731, 86

\bibitem[{{Fowler} {et~al.}(2007){Fowler}, {Niemack}, {Dicker}, {Aboobaker},
  {Ade}, {Battistelli}, {Devlin}, {Fisher}, {Halpern}, {Hargrave}, {Hincks},
  {Kaul}, {Klein}, {Lau}, {Limon}, {Marriage}, {Mauskopf}, {Page}, {Staggs},
  {Swetz}, {Switzer}, {Thornton}, \& {Tucker}}]{Fowler2007}
{Fowler}, J.~W.~{et~al.}, 2007, \ao, 46,
  3444
  
\bibitem[{{Gnedenko}(1943)}]{Gnedenko1943}
{Gnedenko}, B., 1943, Ann. Math., 44, 423

\bibitem[{{Gumbel}(1958)}]{Gumbel1958}
{Gumbel}, E., 1958, {Statistics of Extremes} (Columbia University Press, New
  York (reprinted by Dover, New York in 2004))

\bibitem[{{Harrison} \& {Coles}(2011)}]{Harrison&Coles2011}
{Harrison}, I., {Coles}, P., 2011, preprint (arXiv: 1108.1358)

\bibitem[{{Holz} \& {Perlmutter}(2010)}]{Holz2010}
{Holz}, D.~E., {Perlmutter}, S., 2010, preprint (arXiv: 1004.5349)

\bibitem[{{Hotchkiss}(2011)}]{Hotchkiss2011}
{Hotchkiss}, S., 2011, \jcap, 7, 4

\bibitem[{{Hoyle} {et~al.}(2011){Hoyle}, {Jimenez}, \& {Verde}}]{Hoyle2011}
{Hoyle}, B., {Jimenez}, R., {Verde}, L., 2011, \prd, 83, 103502

\bibitem[{{Hoyle} {et~al.}(2011{\natexlab{b}}){Hoyle}, {Jimenez}, {Verde}, \&
  {Hotchkiss}}]{Hoyle2011b}
{Hoyle}, B., {Jimenez}, R., {Verde}, L., {Hotchkiss}, S., 2011{\natexlab{b}},
  preprint (arXiv: 1108.5458)

\bibitem[{{Jee} {et~al.}(2009){Jee}, {Rosati}, {Ford}, {Dawson}, {Lidman},
  {Perlmutter}, {Demarco}, {Strazzullo}, {Mullis}, {B{\"o}hringer}, \&
  {Fassbender}}]{Jee2009}
{Jee}, M.~J.~{et~al.}, 2009, \apj, 704, 672

\bibitem[{{Jenkinson}(1955)}]{Jenkinson1955}
{Jenkinson}, A.~F., 1955, Quarterly Journal of the Royal Metereological Society,
  81, 158

\bibitem[{{Jimenez} \& {Verde}(2009)}]{Jimenez2009}
{Jimenez}, R., {Verde}, L., 2009, \prd, 80, 127302

\bibitem[{{Komatsu} {et~al.}(1999){Komatsu}, {Kitayama}, {Suto}, {Hattori},
  {Kawabe}, {Matsuo}, {Schindler}, \& {Yoshikawa}}]{Komatsu1999}
{Komatsu}, E., {Kitayama}, T., {Suto}, Y., {Hattori}, M.,
  {Kawabe}, R., {Matsuo}, H., {Schindler}, S., {Yoshikawa}, K., 1999, \apjl, 516, L1

\bibitem[{{Komatsu} {et~al.}(2011){Komatsu}, {Smith}, {Dunkley}, {Bennett},
  {Gold}, {Hinshaw}, {Jarosik}, {Larson}, {Nolta}, {Page}, {Spergel},
  {Halpern}, {Hill}, {Kogut}, {Limon}, {Meyer}, {Odegard}, {Tucker}, {Weiland},
  {Wollack}, \& {Wright}}]{Komatsu2011}
{Komatsu}, E.~{et~al.}, 2011, \apjs, 192, 18

\bibitem[{{Kotz} \& {Nadarajah}(2000)}]{Kotz2000}
{Kotz}, S., {Nadarajah}, S., 2000, {Extreme Value Distributions - Theory and
  Applications} (Imperial College Press, London)
  
\bibitem[{{Laureijs} {et~al.}(2011){Laureijs}, {Amiaux}, {Arduini},
  {Augu{\`e}res}, {Brinchmann}, {Cole}, {Cropper}, {Dabin}, {Duvet}, {Ealet},
  \& et~al.}]{Laureijs2011}
{Laureijs}, R. {et~al.} 2011, preprint (arXiv: 1110.3193)

\bibitem[{{Lewis} \& {Bridle}(2002)}]{Lewis2002}
{Lewis}, A., {Bridle}, S., 2002, \prd, 66, 103511

\bibitem[{{Mantz} {et~al.}(2008){Mantz}, {Allen}, {Ebeling}, \&
  {Rapetti}}]{Mantz2008}
{Mantz}, A., {Allen}, S.~W., {Ebeling}, H., {Rapetti}, D., 2008, \mnras, 387,
  1179

\bibitem[{{Mantz} {et~al.}(2010){Mantz}, {Allen}, {Rapetti}, \&
  {Ebeling}}]{Mantz2010}
{Mantz}, A., {Allen}, S.~W., {Rapetti}, D., {Ebeling}, H., 2010, \mnras, 406,
  1759

\bibitem[{{Markevitch} {et~al.}(2002){Markevitch}, {Gonzalez}, {David},
  {Vikhlinin}, {Murray}, {Forman}, {Jones}, \& {Tucker}}]{Markevitch2002}
{Markevitch}, M., {Gonzalez}, A.~H., {David}, L., {Vikhlinin}, A., {Murray}, S., {Forman}, W., {Jones} C., {Tucker}, W., 2002, \apjl, 567,
  L27
  
\bibitem[{{Marriage} {et~al.}(2011){Marriage}, {Acquaviva}, {Ade}, {Aguirre},
  {Amiri}, {Appel}, {Barrientos}, {Battistelli}, {Bond}, {Brown}, {Burger},
  {Chervenak}, {Das}, {Devlin}, {Dicker}, {Bertrand Doriese}, {Dunkley},
  {D{\"u}nner}, {Essinger-Hileman}, {Fisher}, {Fowler}, {Hajian}, {Halpern},
  {Hasselfield}, {Hern{\'a}ndez-Monteagudo}, {Hilton}, {Hilton}, {Hincks},
  {Hlozek}, {Huffenberger}, {Handel Hughes}, {Hughes}, {Infante}, {Irwin},
  {Baptiste Juin}, {Kaul}, {Klein}, {Kosowsky}, {Lau}, {Limon}, {Lin},
  {Lupton}, {Marsden}, {Martocci}, {Mauskopf}, {Menanteau}, {Moodley},
  {Moseley}, {Netterfield}, {Niemack}, {Nolta}, {Page}, {Parker}, {Partridge},
  {Quintana}, {Reese}, {Reid}, {Sehgal}, {Sherwin}, {Sievers}, {Spergel},
  {Staggs}, {Swetz}, {Switzer}, {Thornton}, {Trac}, {Tucker}, {Warne},
  {Wilson}, {Wollack}, \& {Zhao}}]{Marriage2011}
{Marriage}, T.~A.~{et~al.}, 2011, \apj, 737,
  61
  
  \bibitem[{{Maughan} {et~al.}(2011){Maughan}, {Giles}, {Randall}, {Jones}, \&
  {Forman}}]{Maughan2011}
{Maughan}, B.~J., {Giles}, P.~A., {Randall}, S.~W., {Jones}, C., {Forman},
  W.~R., 2011, preprint (arXiv 1108:1200)
  
\bibitem[{{Melin} {et~al.}(2006){Melin}, {Bartlett}, \&
  {Delabrouille}}]{Melin2006}
{Melin}, J.-B., {Bartlett}, J.~G., {Delabrouille}, J., 2006, \aap, 459, 341

\bibitem[{{Menanteau} {et~al.}(2011){Menanteau}, {Hughes}, {Sifon}, {Hilton},
  {Gonzalez}, {Infante}, {Barrientos}, {Baker}, {Das}, {Devlin}, {Dunkley},
  {Hincks}, {Kosowsky}, {Mardsen}, {Marriage}, {Moodley}, {Niemack}, {Page},
  {Reese}, {Sehgal}, {Sievers}, {Spergel}, {Staggs}, \&
  {Wollack}}]{Menanteau2011}
{Menanteau}, F.~{et~al.}, 2011, preprint (arXiv
  1109.0953)

\bibitem[{{Mortonson} {et~al.}(2011){Mortonson}, {Hu}, \&
  {Huterer}}]{Mortonson2011}
{Mortonson}, M.~J., {Hu}, W., {Huterer}, D., 2011, \prd, 83, 023015

\bibitem[{{Mullis} {et~al.}(2005){Mullis}, {Rosati}, {Lamer}, {B{\"o}hringer},
  {Schwope}, {Schuecker}, \& {Fassbender}}]{Mullis2005}
{Mullis}, C.~R., {Rosati}, P., {Lamer}, G., {B{\"o}hringer}, H.,
  {Schwope}, A., {Schuecker}, P., \& {Fassbender}, R., 2005, \apjl, 623, L85

\bibitem[{{Navarro} {et~al.}(1997){Navarro}, {Frenk}, \& {White}}]{Navarro1997}
{Navarro}, J.~F., {Frenk}, C.~S., {White}, S.~D.~M., 1997, \apj, 490, 493

\bibitem[{{Nord} {et~al.}(2009){Nord}, {Basu}, {Pacaud}, {Ade}, {Bender},
  {Benson}, {Bertoldi}, {Cho}, {Chon}, {Clarke}, {Dobbs}, {Ferrusca},
  {Halverson}, {Holzapfel}, {Horellou}, {Johansson}, {Kennedy}, {Kermish},
  {Kneissl}, {Lanting}, {Lee}, {Lueker}, {Mehl}, {Menten}, {Plagge},
  {Reichardt}, {Richards}, {Schaaf}, {Schwan}, {Spieler}, {Tucker}, {Weiss}, \&
  {Zahn}}]{Nord2009}
{Nord}, M.~ {et~al.}, 2009, \aap, 506, 623

\bibitem[{{Okabe} {et~al.}(2011){Okabe}, {Bourdin}, {Mazzotta}, \&
  {Maurogordato}}]{Okabe2011}
{Okabe}, N., {Bourdin}, H., {Mazzotta}, P., {Maurogordato}, S., 2011, \apj, 741, 116

\bibitem[{{Paranjape} {et~al.}(2011){Paranjape}, {Gordon}, \&
  {Hotchkiss}}]{Paranjape2011}
{Paranjape}, A., {Gordon}, C., {Hotchkiss}, S., 2011, \prd, 84, id:023517

\bibitem[{{Pointecouteau} {et~al.}(1999){Pointecouteau}, {Giard}, {Benoit},
  {D{\'e}sert}, {Aghanim}, {Coron}, {Lamarre}, \&
  {Delabrouille}}]{Pointecouteau1999}
{Pointecouteau}, E., {Giard}, M., {Benoit}, A.,  {D{\'e}sert}, F. X., {Aghanim}, N., {Coron}, N., {Lamarre}, J.~M.,
  {Delabrouille}, J., 1999, \apjl, 519, L115

\bibitem[{{Radovich} {et~al.}(2008){Radovich}, {Puddu}, {Romano}, {Grado}, \&
  {Getman}}]{Radovich2008}
{Radovich}, M., {Puddu}, E., {Romano}, A., {Grado}, A., {Getman}, F., 2008,
  \aap, 487, 55

\bibitem[{{Reiss} \& {Thomas}(2007)}]{Reiss2007}
{Reiss}, R.-D., {Thomas}, M., 2007, {Statistical Analysis of Extreme Values},
  3rd ed. (Birkhauser Verlag, Basel)

\bibitem[{{Rosati} {et~al.}(2009){Rosati}, {Tozzi}, {Gobat}, {Santos},
  {Nonino}, {Demarco}, {Lidman}, {Mullis}, {Strazzullo}, {B{\"o}hringer},
  {Fassbender}, {Dawson}, {Tanaka}, {Jee}, {Ford}, {Lamer}, \&
  {Schwope}}]{Rosati2009}
{Rosati}, P.~ {et~al.}, 2009, \aap, 508, 583

\bibitem[{{Santos} {et~al.}(2011){Santos}, {Fassbender}, {Nastasi},
  {B{\"o}hringer}, {Rosati}, {{\v S}uhada}, {Pierini}, {Nonino},
  {M{\"u}hlegger}, {Quintana}, {Schwope}, {Lamer}, {de Hoon}, \&
  {Strazzullo}}]{Santos2011}
{Santos}, J.~S.~{et~al.}, 2011, \aap, 531,
  L15+

\bibitem[{{Sartoris} {et~al.}(2010){Sartoris}, {Borgani}, {Fedeli},
  {Matarrese}, {Moscardini}, {Rosati}, \& {Weller}}]{Sartoris2010}
{Sartoris}, B., {Borgani}, S., {Fedeli}, C., {Matarrese}, S., {Moscardini}, L., {Rosati}, P., {Weller}, J., 2010, \mnras, 407, 2339

\bibitem[{{Schindler} {et~al.}(1995){Schindler}, {Guzzo}, {Ebeling},
  {Boehringer}, {Chincarini}, {Collins}, {de Grandi}, {Neumann}, {Briel},
  {Shaver}, \& {Vettolani}}]{Schindler1995}
{Schindler}, S.~{et~al.}, 1995, \aap, 299, L9+

\bibitem[{{Schindler} {et~al.}(1997){Schindler}, {Hattori}, {Neumann}, \&
  {Boehringer}}]{Schindler1997}
{Schindler}, S., {Hattori}, M., {Neumann}, D.~M., {Boehringer}, H., 1997,
  \aap, 317, 646

\bibitem[{{Sheth} \& {Tormen}(1999)}]{Sheth&Tormen1999}
{Sheth}, R.~K., {Tormen}, G., 1999, \mnras, 308, 119

\bibitem[{{Squires} {et~al.}(1997){Squires}, {Neumann}, {Kaiser}, {Arnaud},
  {Babul}, {Boehringer}, {Fahlman}, \& {Woods}}]{Squires1997}
{Squires}, G., {Neumann}, D.~M., {Kaiser}, N., {Arnaud}, M.,
  {Babul}, A., {Boehringer}, H., {Fahlman}, G., {Woods}, D., 1997, \apj, 482, 648

\bibitem[{{Sunyaev} \& {Zeldovich}(1980)}]{Sunyaev1980}
{Sunyaev}, R.~A., {Zeldovich}, I.~B., 1980, \araa, 18, 537

\bibitem[{{Sunyaev} \& {Zeldovich}(1972)}]{Sunyaev1972}
{Sunyaev}, R.~A., {Zeldovich}, Y.~B., 1972, Comments on Astrophysics and Space
  Physics, 4, 173
  
\bibitem[{{Tauber, J. A.} {et~al.}(2010){Tauber, J. A.}, {Mandolesi, N.},
  {Puget, J.-L.}, {Banos, T.}, {Bersanelli, M.}, {Bouchet, F. R.}, {Butler, R.
  C.}, {Charra, J.}, {Crone, G.}, {Dodsworth, J.}, {Efstathiou, G.}, {Gispert,
  R.}, {Guyot, G.}, {Gregorio, A.}, {Juillet, J. J.}, {Lamarre, J.-M.},
  {Laureijs, R. J.}, {Lawrence, C. R.}, {N\o{}rgaard-Nielsen, H. U.},
  {Passvogel, T.}, {Reix, J. M.}, {Texier, D.}, {Vibert, L.}, {Zacchei, A.},
  {Ade, P. A. R.}, {Aghanim, N.}, {Aja, B.}, {Alippi, E.}, {Aloy, L.}, {Armand,
  P.}, {Arnaud, M.}, {Arondel, A.}, {Arreola-Villanueva, A.}, {Artal, E.},
  {Artina, E.}, {Arts, A.}, {Ashdown, M.}, {Aumont, J.}, {Azzaro, M.},
  {Bacchetta, A.}, {Baccigalupi, C.}, {Baker, M.}, {Balasini, M.}, {Balbi, A.},
  {Banday, A. J.}, {Barbier, G.}, {Barreiro, R. B.}, {Bartelmann, M.},
  {Battaglia, P.}, {Battaner, E.}, {Benabed, K.}, {Beney, J.-L.}, {Beneyton,
  R.}, {Bennett, K.}, {Benoit, A.}, {Bernard, J.-P.}, {Bhandari, P.}, {Bhatia,
  R.}, {Biggi, M.}, {Biggins, R.}, {Billig, G.}, {Blanc, Y.}, {Blavot, H.},
  {Bock, J. J.}, {Bonaldi, A.}, {Bond, R.}, {Bonis, J.}, {Borders, J.},
  {Borrill, J.}, {Boschini, L.}, {Boulanger, F.}, {Bouvier, J.}, {Bouzit, M.},
  {Bowman, R.}, {Br\'eelle, E.}, {Bradshaw, T.}, {Braghin, M.}, {Bremer, M.},
  {Brienza, D.}, {Broszkiewicz, D.}, {Burigana, C.}, {Burkhalter, M.},
  {Cabella, P.}, {Cafferty, T.}, {Cairola, M.}, {Caminade, S.}, {Camus, P.},
  {Cantalupo, C. M.}, {Cappellini, B.}, {Cardoso, J.-F.}, {Carr, R.},
  {Catalano, A.}, {Cay\'on, L.}, {Cesa, M.}, {Chaigneau, M.}, {Challinor, A.},
  {Chamballu, A.}, {Chambelland, J. P.}, {Charra, M.}, {Chiang, L.-Y.},
  {Chlewicki, G.}, {Christensen, P. R.}, {Church, S.}, {Ciancietta, E.},
  {Cibrario, M.}, {Cizeron, R.}, {Clements, D.}, {Collaudin, B.}, {Colley,
  J.-M.}, {Colombi, S.}, {Colombo, A.}, {Colombo, F.}, {Corre, O.}, {Couchot,
  F.}, {Cougrand, B.}, {Coulais, A.}, {Couzin, P.}, {Crane, B.}, {Crill, B.},
  {Crook, M.}, {Crumb, D.}, {Cuttaia, F.}, {D\"orl, U.}, {da Silva, P.},
  {Daddato, R.}, {Damasio, C.}, {Danese, L.}, {d'Aquino, G.}, {D'Arcangelo,
  O.}, {Dassas, K.}, {Davies, R. D.}, {Davies, W.}, {Davis, R. J.}, {De
  Bernardis, P.}, {de Chambure, D.}, {de Gasperis, G.}, {De la Fuente, M. L.},
  {De Paco, P.}, {De Rosa, A.}, {De Troia, G.}, {De Zotti, G.}, {Dehamme, M.},
  {Delabrouille, J.}, {Delouis, J.-M.}, {D\'esert, F.-X.}, {di Girolamo, G.},
  {Dickinson, C.}, {Doelling, E.}, {Dolag, K.}, {Domken, I.}, {Douspis, M.},
  {Doyle, D.}, {Du, S.}, {Dubruel, D.}, {Dufour, C.}, {Dumesnil, C.}, {Dupac,
  X.}, {Duret, P.}, {Eder, C.}, {Elfving, A.}, {En\ss{}lin, T. A.}, {Eng, P.},
  {English, K.}, {Eriksen, H. K.}, {Estaria, P.}, {Falvella, M. C.}, {Ferrari,
  F.}, {Finelli, F.}, {Fishman, A.}, {Fogliani, S.}, {Foley, S.}, {Fonseca,
  A.}, {Forma, G.}, {Forni, O.}, {Fosalba, P.}, {Fourmond, J.-J.}, {Frailis,
  M.}, {Franceschet, C.}, {Franceschi, E.}, {Fran\c{c}ois, S.}, {Frerking, M.},
  {G\'omez-Re\~nasco, M. F.}, {G\'orski, K. M.}, {Gaier, T. C.}, {Galeotta,
  S.}, {Ganga, K.}, {Garc\'{\i}a L\'azaro, J.}, {Garnica, A.}, {Gaspard, M.},
  {Gavila, E.}, {Giard, M.}, {Giardino, G.}, {Gienger, G.}, {Giraud-Heraud,
  Y.}, {Glorian, J.-M.}, {Griffin, M.}, {Gruppuso, A.}, {Guglielmi, L.},
  {Guichon, D.}, {Guillaume, B.}, {Guillouet, P.}, {Haissinski, J.}, {Hansen,
  F. K.}, {Hardy, J.}, {Harrison, D.}, {Hazell, A.}, {Hechler, M.},
  {Heckenauer, V.}, {Heinzer, D.}, {Hell, R.}, {Henrot-Versill\'e, S.},
  {Hern\'andez-Monteagudo, C.}, {Herranz, D.}, {Herreros, J. M.}, {Hervier,
  V.}, {Heske, A.}, {Heurtel, A.}, {Hildebrandt, S. R.}, {Hills, R.}, {Hivon,
  E.}, {Hobson, M.}, {Hollert, D.}, {Holmes, W.}, {Hornstrup, A.}, {Hovest,
  W.}, {Hoyland, R. J.}, {Huey, G.}, {Huffenberger, K. M.}, {Hughes, N.},
  {Israelsson, U.}, {Jackson, B.}, {Jaffe, A.}, {Jaffe, T. R.}, {Jagemann, T.},
  {Jessen, N. C.}, {Jewell, J.}, {Jones, W.}, {Juvela, M.}, {Kaplan, J.},
  {Karlman, P.}, {Keck, F.}, {Keih\"anen, E.}, {King, M.}, {Kisner, T. S.},
  {Kletzkine, P.}, {Kneissl, R.}, {Knoche, J.}, {Knox, L.}, {Koch, T.},
  {Krassenburg, M.}, {Kurki-Suonio, H.}, {L\"ahteenm\"aki, A.}, {Lagache, G.},
  {Lagorio, E.}, {Lami, P.}, {Lande, J.}, {Lange, A.}, {Langlet, F.}, {Lapini,
  R.}, {Lapolla, M.}, {Lasenby, A.}, {Le Jeune, M.}, {Leahy, J. P.}, {Lefebvre,
  M.}, {Legrand, F.}, {Le Meur, G.}, {Leonardi, R.}, {Leriche, B.}, {Leroy,
  C.}, {Leutenegger, P.}, {Levin, S. M.}, {Lilje, P. B.}, {Lindensmith, C.},
  {Linden-V\o{}rnle, M.}, {Loc, A.}, {Longval, Y.}, {Lubin, P. M.}, {Luchik,
  T.}, {Luthold, I.}, {Macias-Perez, J. F.}, {Maciaszek, T.}, {MacTavish, C.},
  {Madden, S.}, {Maffei, B.}, {Magneville, C.}, {Maino, D.}, {Mambretti, A.},
  {Mansoux, B.}, {Marchioro, D.}, {Maris, M.}, {Marliani, F.}, {Marrucho,
  J.-C.}, {Mart\'{\i}-Canales, J.}, {Mart\'{\i}nez-Gonz\'alez, E.},
  {Mart\'{\i}n-Polegre, A.}, {Martin, P.}, {Marty, C.}, {Marty, W.}, {Masi,
  S.}, {Massardi, M.}, {Matarrese, S.}, {Matthai, F.}, {Mazzotta, P.},
  {McDonald, A.}, {McGrath, P.}, {Mediavilla, A.}, {Meinhold, P. R.}, {M\'elin,
  J.-B.}, {Melot, F.}, {Mendes, L.}, {Mennella, A.}, {Mervier, C.}, {Meslier,
  L.}, {Miccolis, M.}, {Miville-Deschenes, M.-A.}, {Moneti, A.}, {Montet, D.},
  {Montier, L.}, {Mora, J.}, {Morgante, G.}, {Morigi, G.}, {Morinaud, G.},
  {Morisset, N.}, {Mortlock, D.}, {Mottet, S.}, {Mulder, J.}, {Munshi, D.},
  {Murphy, A.}, {Murphy, P.}, {Musi, P.}, {Narbonne, J.}, {Naselsky, P.},
  {Nash, A.}, {Nati, F.}, {Natoli, P.}, {Netterfield, B.}, {Newell, J.},
  {Nexon, M.}, {Nicolas, C.}, {Nielsen, P. H.}, {Ninane, N.}, {Noviello, F.},
  {Novikov, D.}, {Novikov, I.}, {O'Dwyer, I. J.}, {Oldeman, P.}, {Olivier, P.},
  {Ouchet, L.}, {Oxborrow, C. A.}, {P\'erez-Cuevas, L.}, {Pagan, L.}, {Paine,
  C.}, {Pajot, F.}, {Paladini, R.}, {Pancher, F.}, {Panh, J.}, {Parks, G.},
  {Parnaudeau, P.}, {Partridge, B.}, {Parvin, B.}, {Pascual, J. P.}, {Pasian,
  F.}, {Pearson, D. P.}, {Pearson, T.}, {Pecora, M.}, {Perdereau, O.},
  {Perotto, L.}, {Perrotta, F.}, {Piacentini, F.}, {Piat, M.}, {Pierpaoli, E.},
  {Piersanti, O.}, {Plaige, E.}, {Plaszczynski, S.}, {Platania, P.},
  {Pointecouteau, E.}, {Polenta, G.}, {Ponthieu, N.}, {Popa, L.}, {Poulleau,
  G.}, {Poutanen, T.}, {Pr\'ezeau, G.}, {Pradell, L.}, {Prina, M.}, {Prunet,
  S.}, {Rachen, J. P.}, {Rambaud, D.}, {Rame, F.}, {Rasmussen, I.},
  {Rautakoski, J.}, {Reach, W. T.}, {Rebolo, R.}, {Reinecke, M.}, {Reiter, J.},
  {Renault, C.}, {Ricciardi, S.}, {Rideau, P.}, {Riller, T.}, {Ristorcelli,
  I.}, {Riti, J. B.}, {Rocha, G.}, {Roche, Y.}, {Pons, R.}, {Rohlfs, R.},
  {Romero, D.}, {Roose, S.}, {Rosset, C.}, {Rouberol, S.}, {Rowan-Robinson,
  M.}, {Rubi\~no-Mart\'{\i}n, J. A.}, {Rusconi, P.}, {Rusholme, B.}, {Salama,
  M.}, {Salerno, E.}, {Sandri, M.}, {Santos, D.}, {Sanz, J. L.}, {Sauter, L.},
  {Sauvage, F.}, {Savini, G.}, {Schmelzel, M.}, {Schnorhk, A.}, {Schwarz, W.},
  {Scott, D.}, {Seiffert, M. D.}, {Shellard, P.}, {Shih, C.}, {Sias, M.},
  {Silk, J. I.}, {Silvestri, R.}, {Sippel, R.}, {Smoot, G. F.}, {Starck,
  J.-L.}, {Stassi, P.}, {Sternberg, J.}, {Stivoli, F.}, {Stolyarov, V.},
  {Stompor, R.}, {Stringhetti, L.}, {Strommen, D.}, {Stute, T.}, {Sudiwala,
  R.}, {Sugimura, R.}, {Sunyaev, R.}, {Sygnet, J.-F.}, {T\"urler, M.}, {Taddei,
  E.}, {Tallon, J.}, {Tamiatto, C.}, {Taurigna, M.}, {Taylor, D.}, {Terenzi,
  L.}, {Thuerey, S.}, {Tillis, J.}, {Tofani, G.}, {Toffolatti, L.}, {Tommasi,
  E.}, {Tomasi, M.}, {Tonazzini, E.}, {Torre, J.-P.}, {Tosti, S.}, {Touze, F.},
  {Tristram, M.}, {Tuovinen, J.}, {Tuttlebee, M.}, {Umana, G.}, {Valenziano,
  L.}, {Vall\'ee, D.}, {van der Vlis, M.}, {Van Leeuwen, F.}, {Vanel, J.-C.},
  {Van-Tent, B.}, {Varis, J.}, {Vassallo, E.}, {Vescovi, C.}, {Vezzu, F.},
  {Vibert, D.}, {Vielva, P.}, {Vierra, J.}, {Villa, F.}, {Vittorio, N.},
  {Vuerli, C.}, {Wade, L. A.}, {Walker, A. R.}, {Wandelt, B. D.}, {Watson, C.},
  {Werner, D.}, {White, M.}, {White, S. D. M.}, {Wilkinson, A.}, {Wilson, P.},
  {Woodcraft, A.}, {Yoffo, B.}, {Yun, M.}, {Yurchenko, V.}, {Yvon, D.}, {Zhang,
  B.}, {Zimmermann, O.}, {Zonca, A.}, \& {Zorita, D.}}]{Tauber2010}
{Tauber}, J. A. {et~al.}, 2010, \aap, 520, A1  

\bibitem[{{Tinker} {et~al.}(2008){Tinker}, {Kravtsov}, {Klypin}, {Abazajian},
  {Warren}, {Yepes}, {Gottl{\"o}ber}, \& {Holz}}]{Tinker2008}
{Tinker}, J., {Kravtsov}, A.~V., {Klypin}, A., {Abazajian}, K.,
  {Warren}, M., {Yepes}, G., {Gottl{\"o}ber}, S., {Holz}, D.~E., 2008, \apj, 688, 709
  
\bibitem[{{Tucker} {et~al.}(1998){Tucker}, {Blanco}, {Rappoport}, {David},
  {Fabricant}, {Falco}, {Forman}, {Dressler}, \& {Ramella}}]{Tucker1998}
{Tucker}, W. {et~al.}, 1998, \apjl, 496, L5+

\bibitem[{{Umetsu} {et~al.}(2011){Umetsu}, {Broadhurst}, {Zitrin},
  {Medezinski}, \& {Hsu}}]{Umetsu2011}
{Umetsu}, K., {Broadhurst}, T., {Zitrin}, A., {Medezinski}, E., {Hsu}, L.-Y.,
  2011, \apj, 729, 127

\bibitem[{{Vanderlinde} {et~al.}(2010){Vanderlinde}, {Crawford}, {de Haan},
  {Dudley}, {Shaw}, {Ade}, {Aird}, {Benson}, {Bleem}, {Brodwin}, {Carlstrom},
  {Chang}, {Crites}, {Desai}, {Dobbs}, {Foley}, {George}, {Gladders}, {Hall},
  {Halverson}, {High}, {Holder}, {Holzapfel}, {Hrubes}, {Joy}, {Keisler},
  {Knox}, {Lee}, {Leitch}, {Loehr}, {Lueker}, {Marrone}, {McMahon}, {Mehl},
  {Meyer}, {Mohr}, {Montroy}, {Ngeow}, {Padin}, {Plagge}, {Pryke}, {Reichardt},
  {Rest}, {Ruel}, {Ruhl}, {Schaffer}, {Shirokoff}, {Song}, {Spieler},
  {Stalder}, {Staniszewski}, {Stark}, {Stubbs}, {van Engelen}, {Vieira},
  {Williamson}, {Yang}, {Zahn}, \& {Zenteno}}]{Vanderlinde2010}
{Vanderlinde}, K.~{et~al.}, 2010, \apj, 722,
  1180

\bibitem[{{von Mises}(1954)}]{vonmises1954}
{von Mises}, R., 1954, Americ. Math. Soc, Volume II, 271

\bibitem[{{Waizmann} {et~al.}(2011){Waizmann}, {Ettori}, \&
  {Moscardini}}]{Waizmann2011}
{Waizmann}, J.-C., {Ettori}, S., {Moscardini}, L., 2011, MNRAS, in print, doi: 10.1111/j.1365-2966.2011.19496.x

\bibitem[{{Williamson} {et~al.}(2011){Williamson}, {Benson}, {High},
  {Vanderlinde}, {Ade}, {Aird}, {Andersson}, {Armstrong}, {Ashby}, {Bautz},
  {Bazin}, {Bertin}, {Bleem}, {Bonamente}, {Brodwin}, {Carlstrom}, {Chang},
  {Chapman}, {Clocchiatti}, {Crawford}, {Crites}, {de Haan}, {Desai}, {Dobbs},
  {Dudley}, {Fazio}, {Foley}, {Forman}, {Garmire}, {George}, {Gladders},
  {Gonzalez}, {Halverson}, {Holder}, {Holzapfel}, {Hoover}, {Hrubes}, {Jones},
  {Joy}, {Keisler}, {Knox}, {Lee}, {Leitch}, {Lueker}, {Luong-Van}, {Marrone},
  {McMahon}, {Mehl}, {Meyer}, {Mohr}, {Montroy}, {Murray}, {Padin}, {Plagge},
  {Pryke}, {Reichardt}, {Rest}, {Ruel}, {Ruhl}, {Saliwanchik}, {Saro},
  {Schaffer}, {Shaw}, {Shirokoff}, {Song}, {Spieler}, {Stalder}, {Stanford},
  {Staniszewski}, {Stark}, {Story}, {Stubbs}, {Vieira}, {Vikhlinin}, \&
  {Zenteno}}]{Williamson2011}
{Williamson}, R.~{et~al.}, 2011, \apj, 738, 139
  
\bibitem[{{Zhao} {et~al.}(2009){Zhao}, {Jing}, {Mo}, \&
  {B{\"o}rner}}]{Zhao2009}
{Zhao}, D.~H., {Jing}, Y.~P., {Mo}, H.~J., {B{\"o}rner}, G., 2009, \apj, 707,
  354  

\end{thebibliography}

\end{document}